\documentclass[sn-mathphys,Numbered]{sn-jnl}% Math and Physical Sciences Reference Style
\usepackage{graphicx}%
\usepackage{multirow}%
\usepackage{amsmath,amssymb,amsfonts}%
\usepackage{amsthm}%
\usepackage{mathrsfs}%
\usepackage[title]{appendix}%
\usepackage{textcomp}%
\usepackage{manyfoot}%
\usepackage{booktabs}%
\usepackage{listings}%
\usepackage{amsfonts,amssymb,stmaryrd,latexsym,amsmath,braket}
\usepackage{graphicx,subfigure}
\usepackage{comment}
\usepackage{times}
\usepackage{slashed}
\usepackage[dvipsnames,usenames]{xcolor}
\usepackage{lipsum,adjustbox}
\usepackage[capitalise]{cleveref}
\usepackage{tikz}
\usetikzlibrary{quantikz}
\usepackage{bm} % for bold math symbols
\begin{document}

\title{Controlled Gate Networks: Theory and Application to Eigenvalue Estimation}

\author[1]{Max Bee-Lindgren}
\affil[1]{School of Physics, Georgia Institute of Technology, Atlanta, GA 30332, USA}
\author[2]{Zhengrong Qian}
\affil[2]{Facility for Rare Isotope Beams and Department of Physics and
Astronomy, Michigan State University, MI 48824, USA}
\author[3]{Matthew DeCross}
\affil[3]{Quantinuum, 303 S. Technology Ct., Broomfield, Colorado 80021, USA}
\author[3]{Natalie C. Brown}
%\affiliation{Quantinuum, 303 S. Technology Ct., Broomfield, Colorado 80021, USA}
\author[3]{Christopher N. Gilbreth}
%\affiliation{Quantinuum, 303 S. Technology Ct., Broomfield, Colorado 80021, USA}
\author[2]{Jacob Watkins}
\author[2]{Xilin Zhang}
\author[2]{Dean Lee}
%\affiliation{Facility for Rare Isotope Beams and Department of Physics and Astronomy, Michigan State University, MI 48824, USA}
\abstract{
We introduce a new scheme for quantum circuit design called controlled gate networks.  Rather than trying to reduce the complexity of individual unitary operations, the new strategy is to toggle between all of the unitary operations needed with the fewest number of gates.   We present the general theory of controlled gate networks and show that, under quite general conditions, it can significantly reduce the number of two-qubit gates needed to produce linear combinations of unitary operators.  The first example we consider is a variational subspace calculation for a two-qubit system.  The second example is estimating the eigenvalues of a two-qubit Hamiltonian via the rodeo algorithm \cite{Choi:2020pdg} using operators that we call controlled reversal gates.  We use the Quantinuum H1-2 and IBM Perth devices to realize the quantum circuits.  The third example is the application of controlled gate networks to the controlled time evolution of a free nucleon on a three-dimensional lattice.  For all of the examples, we show very substantial reductions in the number of two-qubit gates required. Our work demonstrates that controlled gate networks are a useful tool for reducing gate complexity in quantum algorithms for quantum many-body problems such as those relevant to nuclear physics.}

\maketitle

\section*{Introduction}
In low-energy nuclear physics, many different {\it ab initio} methods are being used for calculations of nuclear structure and reactions using classical computing~\cite{Navratil:2003ef,Maris:2008ax,Barrett:2013nh,Barbieri:2016uib,Lonardoni:2017hgs,Wirth:2017bpw,Piarulli:2017dwd,Lonardoni:2018nob,Tichai:2018vjc,Roggero:2018hrn,Roggero:2020qoz,Hupin:2018biv,Sun:2018fmu,Dytrych:2018vkl,Smirnova:2019yiq,Holt:2019gmc,Dawkins:2019vcr,Idini:2019hkq,Yao:2019rck,Dreyfuss:2020lss,Tichai:2020dna,Stroberg:2019mxo,Demol:2020mzd,Jiang:2020the,Raghavan:2020bze,Sobczyk:2021dwm,Arthuis:2022ixv,Shen:2022bak,Elhatisari:2022zrb}.  While there has been much progress in recent years, there is substantial opportunity for a new computational approach such as quantum computing to address some of the many difficult challenges that remain in preparing eigenstates for the nuclear many-body problem.  In particular, several of the classical computational methods have difficulties in accurately probing the spectrum of states at energies significantly higher than that of the ground state.  Several recent reviews on nuclear many-body physics and quantum computing algorithms can be found in Ref.~\cite{Ayral:2023ron,Lee:2023izc,Garcia-Ramos:2023rtd,Bauer:2023qgm}.
 
 Many quantum circuits involve combinations of unitary operators executed in sequence, with logical control directed by one or more auxiliary registers~\cite{10.5555/2481569.2481570,berry2014exponential,low2019hamiltonian,gilyen2019quantum}. In order to minimize circuit complexity, many promising techniques and strategies have been developed \cite{Khatri2019quantumassisted,Kamaka2020QuantumTO,2020PhRvA.102f2612Y,Nash_2020,9951320,Dobbs:2022ghc,Weiden:2023nkc}.  For circuits with controlled unitary operations, the conventional approach is to reduce the gate complexity of each controlled unitary operator as much as possible. During the transpilation stage, simplifications between circuit blocks may be found and exploited. However, this alone does not suggest how to design circuits intentionally such that cancellations are common.   
 
  Here we present a different approach in which all the unitary operators needed are viewed as a connected network, and the transformations between the operators are optimized. We define a controlled gate network as a set of controlled unitary operators that are transformed into each other by the controlled inclusion or exclusion of operators that we call transformation gates.  
 
 General strategies for reducing circuit complexity are based on local reductions of circuit substructures that are known to be reducible to simpler structures \cite{chen2021quantum,iten2022exact,ge2024quantum,karuppasamy2025comprehensive}.  Some examples are gate fusion, gate cancellation, and gate commutation. Transpiling software makes use of these local optimizations to create circuits with fewer gates but the same functionality. In contrast, controlled gate networks focus on identifying higher-level structural relationships between sets of controlled unitary operators.  These relationships may involve many gates and involve complex structures that are difficult for local optimization methods to discover.  While a different strategy, it is important to note that controlled gate networks can be used together with local optimization methods to find new pathways to achieve greater reductions in circuit complexity.
 
\begin{figure}
    \centering
    \includegraphics[width=10cm,trim=4 4 4 4,clip]{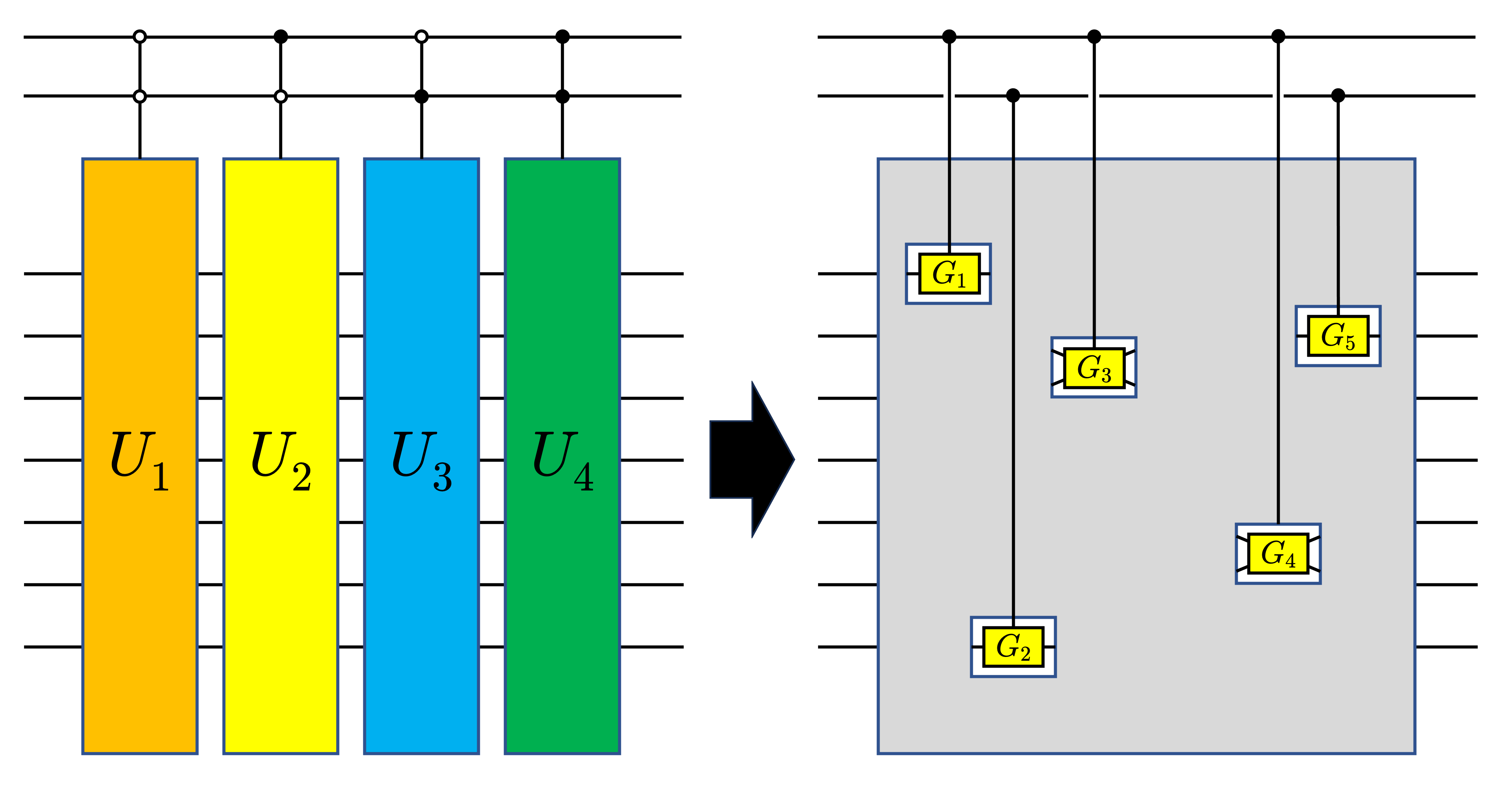}
\caption{Example of a controlled gate network.  The left side shows a circuit using two ancilla qubits to toggle between four possible unitary operators, $U_1, U_2, U_3, U_4$.  The filled (open) dot means that the controlled operator is actuated when the attached ancilla is in the $\ket{1}$ $(\ket{0})$ state.  The right side achieves the same results using a controlled gate network.  Each of the transformation gates $G_i$ act on a small number of qubits.  $U_1$ corresponds to no $G_i$ actuated, $U_2$ corresponds to actuating $\{G_1,G_3,G_4\}$, $U_3$ corresponds to $\{G_2,G_5\}$, and $U_4$ corresponds to $\{G_1,G_2,G_3,G_4,G_5\}$.}
\label{fig:controlled_gate_network}
\end{figure}

 An example of a controlled gate network is illustrated in Fig.~\ref{fig:controlled_gate_network}.  The left side shows a circuit using two ancilla qubits to toggle between four possible unitary operators, $U_1, U_2, U_3, U_4$.  The right side achieves the same results using a controlled gate network.  Each of the transformation gates $G_i$ act on a small number of qubits, and the ancilla qubits control the transformation gates.  Further details are given in the figure caption.
 
Controlled gate networks are useful for quantum algorithms that use controlled operator structures.  Examples include linear combinations of unitary operators for Hamiltonian simulation and quantum signal processing \cite{10.5555/2481569.2481570,low2019hamiltonian}, phase estimation \cite{Kitaev:1995qy,Abrams:1999,Cleve:1998,Svore:2013}, imaginary time evolution \cite{motta2020determining,Turro:2021vbk}, and variational subspace methods \cite{2019arXiv190908925P,2022arXiv220910571F,2022arXiv221114854Y,2022PhRvA.105b2417C,Mejuto-Zaera:2023ier}. Controlled unitary operators that are similar in operator structure are well-suited for controlled gate networks, as then the transformation gates can be quite simple.

In what follows, we outline the general theory of controlled gate networks and show how this approach minimizes the number of two-qubit gates required to construct linear combinations of unitary operators. We then provide three illustrative examples. The first involves a variational subspace calculation for a two-qubit system. The second demonstrates eigenvalue estimation of a two-qubit Hamiltonian using the rodeo algorithm \cite{Choi:2020pdg} with a particular type of controlled gate network known as a controlled reversal gate. The third example applies controlled gate networks to simulate the controlled time evolution of a free nucleon on a three-dimensional lattice.

%The purpose of this article is twofold.  The first is to introduce the concept of a controlled gate network and show how it can reduce the number of gates in circuits when multiple, simply related control operations are required.  The second is to show the implementation of controlled gate networks for variational subspace methods and the rodeo algorithm.
 
\section{General theory of controlled gate networks}
Any observable measured on a quantum computer corresponds to the expectation of some Hermitian observable $O$ for some quantum state $\ket{\Psi}$.  We prepare $\ket{\Psi}$ by starting from some simple initial state $\ket{\Phi}$ and constructing some linear combination of unitary operations $W_j$ acting on the initial state,
\begin{equation}
    \ket{\Psi}=\sum_j c_j W_j \ket{\Phi}.
\end{equation}
The expectation value of $O$ is then 
\begin{equation}
    \sum_j \sum_{j'} c^*_j c_{j'} \braket{\Phi|W^\dagger_j O W_{j'} |\Phi}.
\end{equation}
For linear combinations of a small number of unitary operators, we can determine $\braket{\Phi|W^\dagger_j O W_{j'} |\Phi}$ by computing the expectation values of $O$ with only pairwise linear combinations of $W_j$ and $W_{j'}$.  For linear combinations of a large number of unitary operators, however, constructing the full linear combination is often more computational efficient.  In the following, we consider each of the two applications, pairwise linear combinations and total linear combinations.

\begin{figure}
    \centering
\includegraphics[width=6cm]{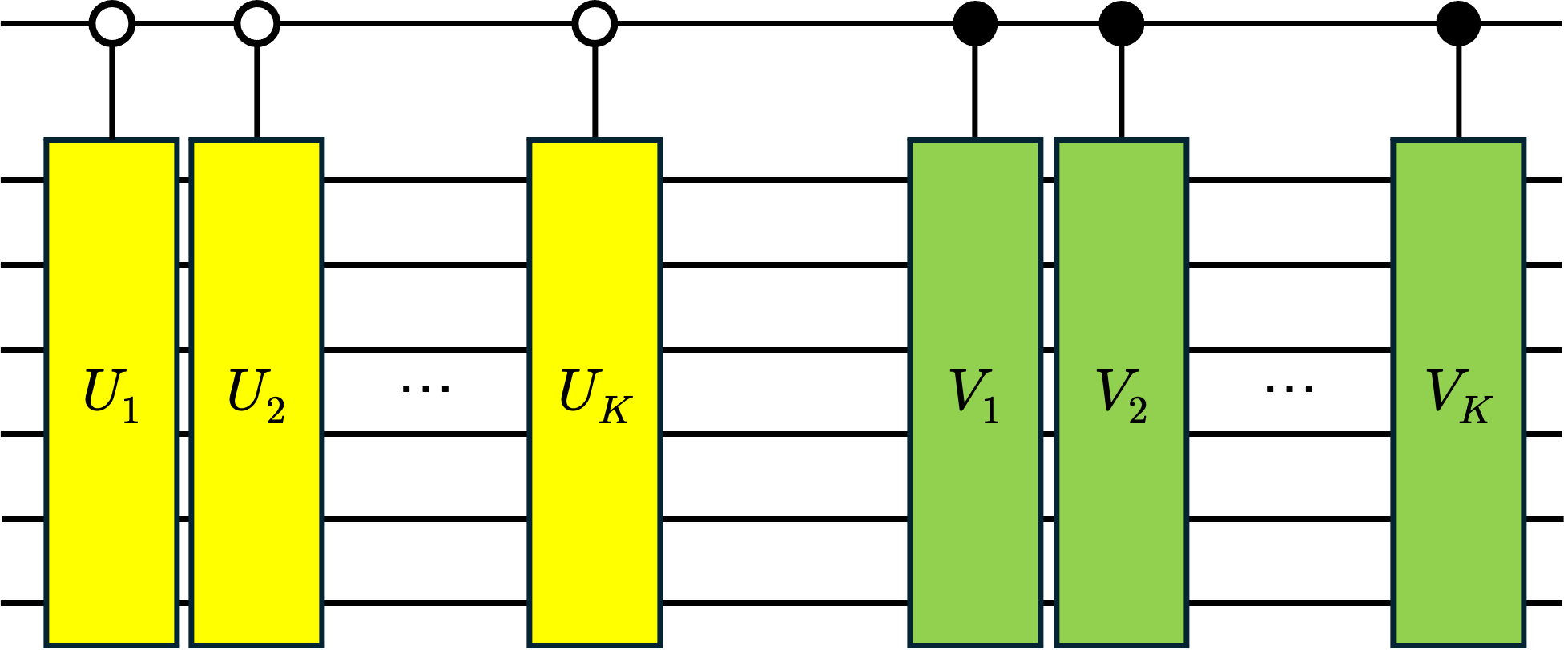}
\caption{Standard approach for producing linear combinations of $U$ and $V$ with one ancilla qubit by controlling $U_1, U_2, \cdots, U_K$ with the $0$ state and controlling $V_1, V_2, \cdots, V_K$ with the $1$ state.}
\label{fig:regular}
\end{figure}

We start with the problem of constructing linear combinations of two unitary operators, which we call $U$ and $V$.  Let us decompose $U$ and $V$ as a product of $K$ simpler unitary operators
\begin{align}
U = U_1 U_2 \cdots U_K \nonumber \\
V = V_1 V_2 \cdots V_K.
\end{align}
We can produce any linear combination of $U$ and $V$ with one ancilla qubit by controlling $U_1, U_2, \cdots, U_K$ with the $0$ state and controlling $V_1, V_2, \cdots, V_K$ with the $1$ state.  This is illustrated in Fig.~\ref{fig:regular}.  This standard approach does not take advantage of similarities in the structures of $U$ and $V$ and performs control operations for each one independent of the other.   

\begin{figure}
    \centering
\includegraphics[width=6cm]{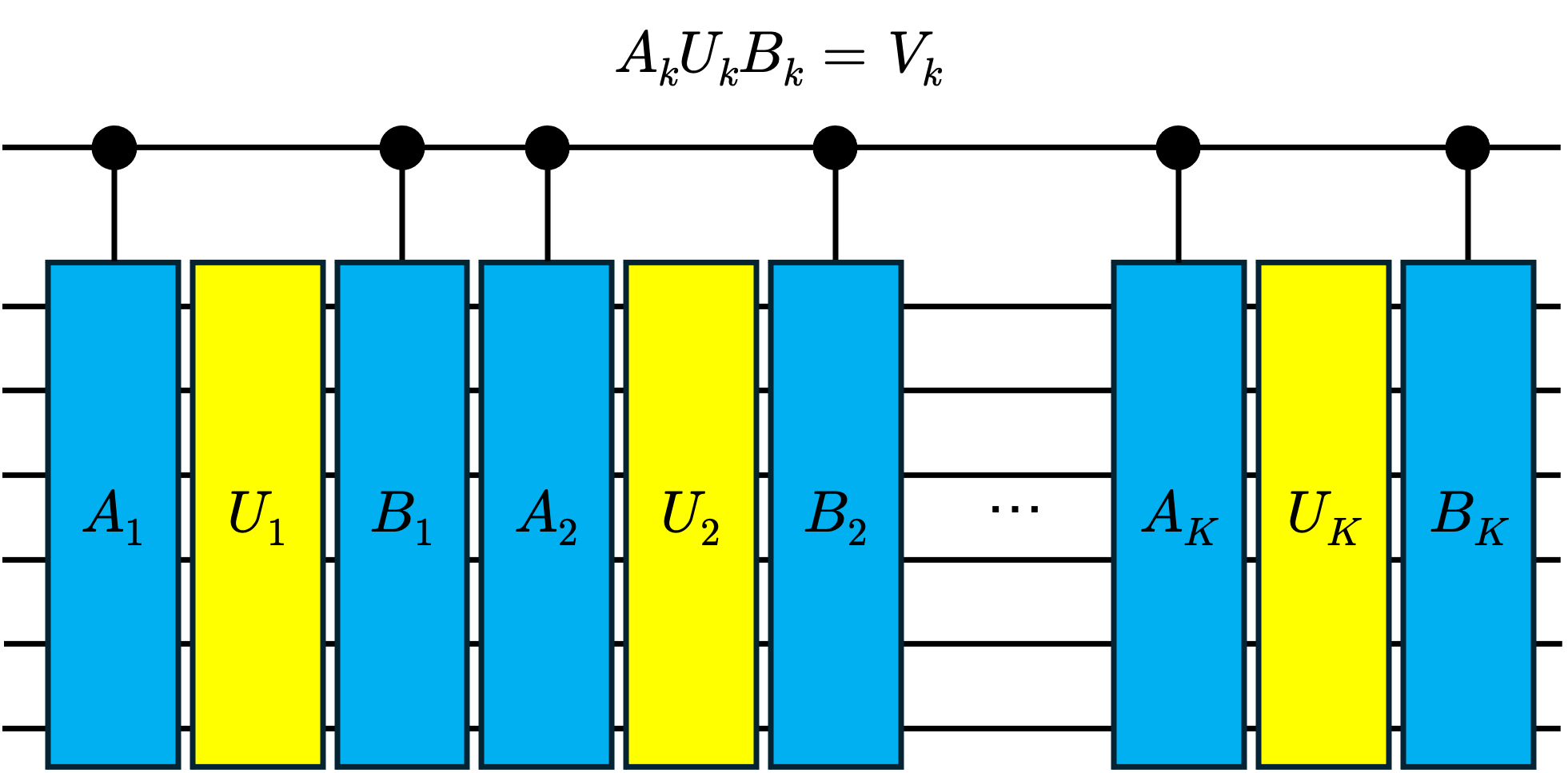}
\caption{Controlled gate network approach for producing linear combinations of $U$ and $V$ with one ancilla qubit.  We control each $A_k$ and $B_k$ with the 1 state.}
\label{fig:controlled}
\end{figure}

We can also generate the same linear combination using a controlled gate network.  For each $U_k$ and $V_k$, we find unitary operators $A_k$ and $B_k$ such that $A_k U_k B_k = V_k$.  These unitary operators are the transformation gates in our controlled gate network.  $A_k$ can be any unitary operator, so long as $B_k = U^\dagger_k A^\dagger_k V_k$.  For the special case that $A_k$ is the identity, we see that $B_k = U^\dagger_k V_k$. The general strategy is to choose $A_k$ so that both $A_k$ and $B_k$ are simple operators that can be controlled efficiently with an ancilla qubit.  We can produce general linear combinations of $U$ and $V$ with a single ancilla qubit by controlling each $A_k$ and $B_k$. This is illustrated in Fig.~\ref{fig:controlled}.  

In order to quantify the computational cost of various quantum circuits, we count the number of CNOT gates needed when decomposed into CNOT gates and arbitrary single-qubit gates.  Let $\nu$ be the minimum number of CNOT gates needed for a given circuit.  We note that for any unitary proportional to the identity, $e^{i\theta} I$, we have $\nu(e^{i\theta} I) = 0$ since it requires only a phase gate on the ancilla qubit.  

Consider any unitary operator $W$ that is not proportional to the identity.  Let $C(W)$ be the controlled unitary operator associated with $W$.  When the information is relevant, we write $C^{(0)}(W)$ or $C^{(1)}(W)$ to indicate that the controlling qubit is conditioned on the $0$ state or $1$ state respectively.  It follows that 
\begin{equation}
    \nu(C(W)) \ge \nu(W)+1.
    \label{eq:CW_vs_W}
\end{equation}  
The proof is as follows.  Since $W$ is not proportional to the identity, we must have at least one CNOT gate connecting the control qubit $q$ to the qubits comprising $W$.  Suppose now that $\nu(C^{(i)}(W)) \le \nu(W)$, then the projection of $C^{(i)}(W)$ onto the state $i$ for $q$ gives a representation of $W$ with only $\nu(W)-1$ CNOT gates or less.  This contradiction proves that $\nu(C(W)) \ge \nu(W)+1$.  By induction, we have also proved that if we control $W$ with $n$ qubits, then the required number of CNOT gates is greater than or equal to $\nu(C(W))+n$.

In the standard approach, the number of CNOT gates to control $U_k$ with state 0 and control $V_k$ with state 1 is then $\nu(C(U_k)) + \nu(C(V_k))$. Suppose that we now make a simple choice of setting $A_k$ equal to the identity.  Then the CNOT gate count for the controlled gate network at step $k$ is $\nu(U_k) + \nu(C(U^\dagger_k V_k))$.  Let us now consider several different cases. If $U_k$ and $V_k$ are equal up to a phase, $V_k=e^{i\theta}U_k$, then $\nu(C(U^\dagger_k V_k)) = 0$.  For this case, the controlled gate network reduces the CNOT count by an amount $\nu(C(V_k))+\nu(C(U_k))-\nu(U_k)$. 

Next, we consider the case where $\nu(C(U^\dagger_k V_k)) \le \nu(C(V_k))$ and $U_k$ are not proportional to the identity.  These conditions are quite easily satisfied if $U_k$ and $V_k$ act on the same set of qubits and $V_k$ does not have a special form that makes $\nu(C(V_k))$ smaller than $\nu(C(U^\dagger_kV_k))$.  Using Eq.~\eqref{eq:CW_vs_W}, we find that
\begin{equation}
    \nu(U_k) + \nu(C(U^\dagger_k V_k)) \le \nu(C(U_k)) + \nu(C(V_k)) - 1.
    \label{eq:CNOTreduction}
\end{equation}
We conclude that the controlled gate network at step $k$ reduces the number of CNOT gates by at least $1$.  In many cases, the reduction is significantly more than $1$, but this is a rigorous argument showing that the reduction is at least $1$ as long as $\nu(C(U^\dagger_k V_k)) \le \nu(C(V_k))$. In order to maximize the reduction in the number of CNOT gates, one can reverse the roles of $U$ and $V$ when $\sum_{k=1}^K{\nu(C(V_k))}$ is less than $\sum_{k=1}^K{\nu(C(U_k))}$.  If $U$ and $V$ are similar to each other in terms of gate structure, then it is fairly easy to satisfy these conditions for several of the $k=1,2,\cdots,K$ steps.  The total reduction in the number of CNOT gates can therefore be quite large.  

This demonstrates the general utility of controlled gate networks in reducing the number of CNOT gates needed when producing linear combinations of two unitary operators $U$ and $V$. By decomposing $U$ and $V$ into the products of operators $U_k$ and $V_k$, the controlled gate network exploits similarities in the gate structures of $U_k$ and $V_k$. In our analysis, we did not consider the additional degree of freedom available in optimizing $A_k$ and using the more general relation $B_k = U^\dagger_k A^\dagger_k V_k$.  As we show later for the case of controlled reversal gates, optimizing the choice for $A_k$ can produce even greater reductions in the number of CNOT gates. 

\begin{figure}
    \centering
\includegraphics[width=6cm]{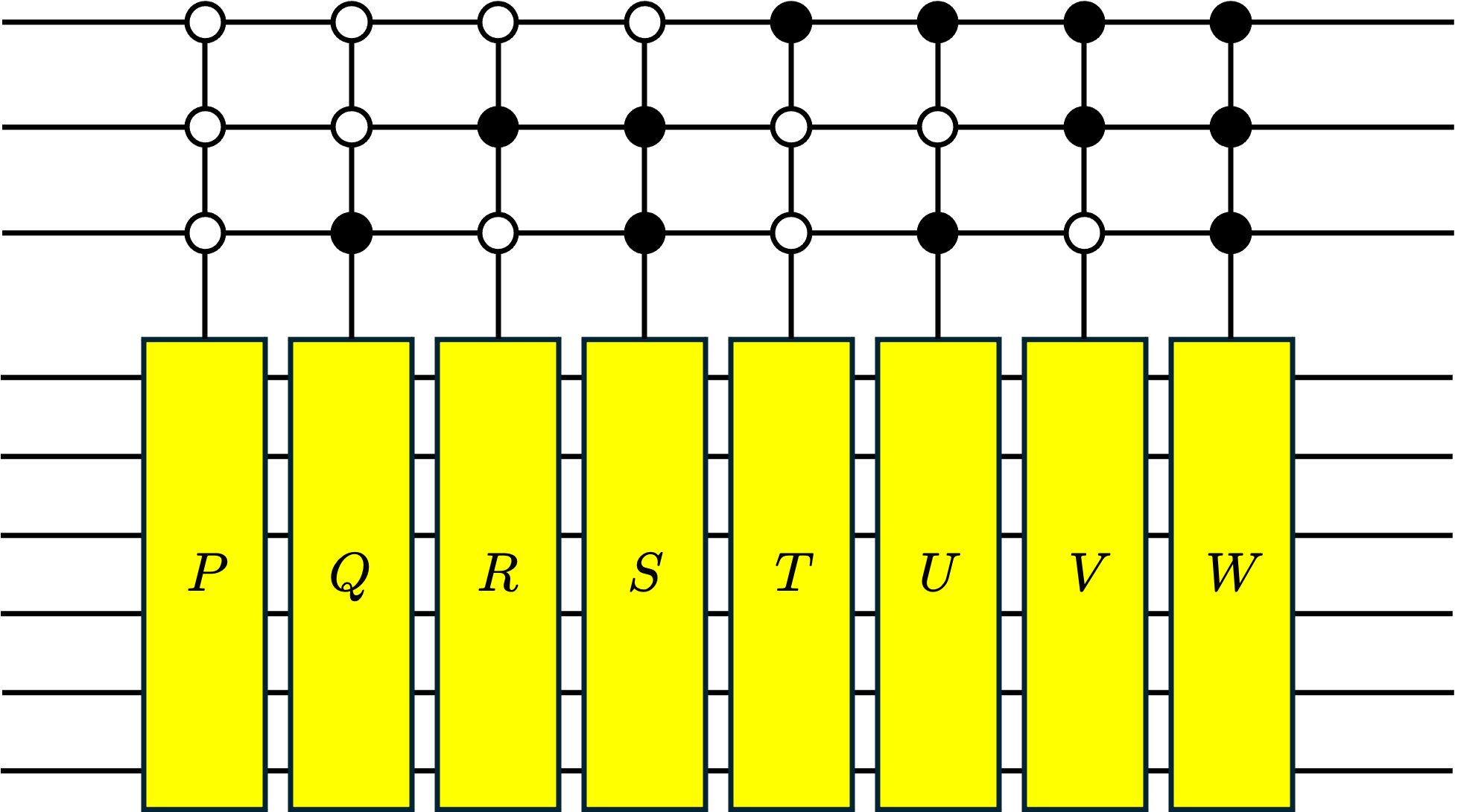}
\caption{Standard approach for producing a linear combination of eight unitary operators using three ancilla qubits.}
\label{fig:LCU_regular}
\end{figure}

We now consider the problem of creating a linear combination of more than two unitary operators.  Suppose that we want to produce a linear combination of $N = 2^n$ unitary operators using $n>1$ ancilla qubits. The standard approach is illustrated in Fig.~\ref{fig:LCU_regular} for $N = 8$ unitary operators and $n = 3$ ancilla qubits.  If we read the control states from top to bottom as binary digits, we have all possible binary digits for the integers from $0$ to $2^n-1$.  We will call these control-state binary digits.  We can write each unitary operator as products of simpler unitary operators,
\begin{align}
    & P = P_1 P_2 \cdots P_K, \; Q = Q_1 Q_2 \cdots Q_K, \nonumber \\
    & R = R_1 R_2 \cdots R_K, \; S = S_1 S_2 \cdots S_K, \nonumber \\
    & T = T_1 T_2 \cdots T_K, \; U = U_1 U_2 \cdots U_K, \nonumber \\
    & V = V_1 V_2 \cdots V_K, \; W = W_1 W_2 \cdots W_K \nonumber, 
    \label{eq:steps}
\end{align}
and perform the control operations at each step $k=1, \cdots, K$.

\begin{figure}
    \centering
\includegraphics[width=6cm]{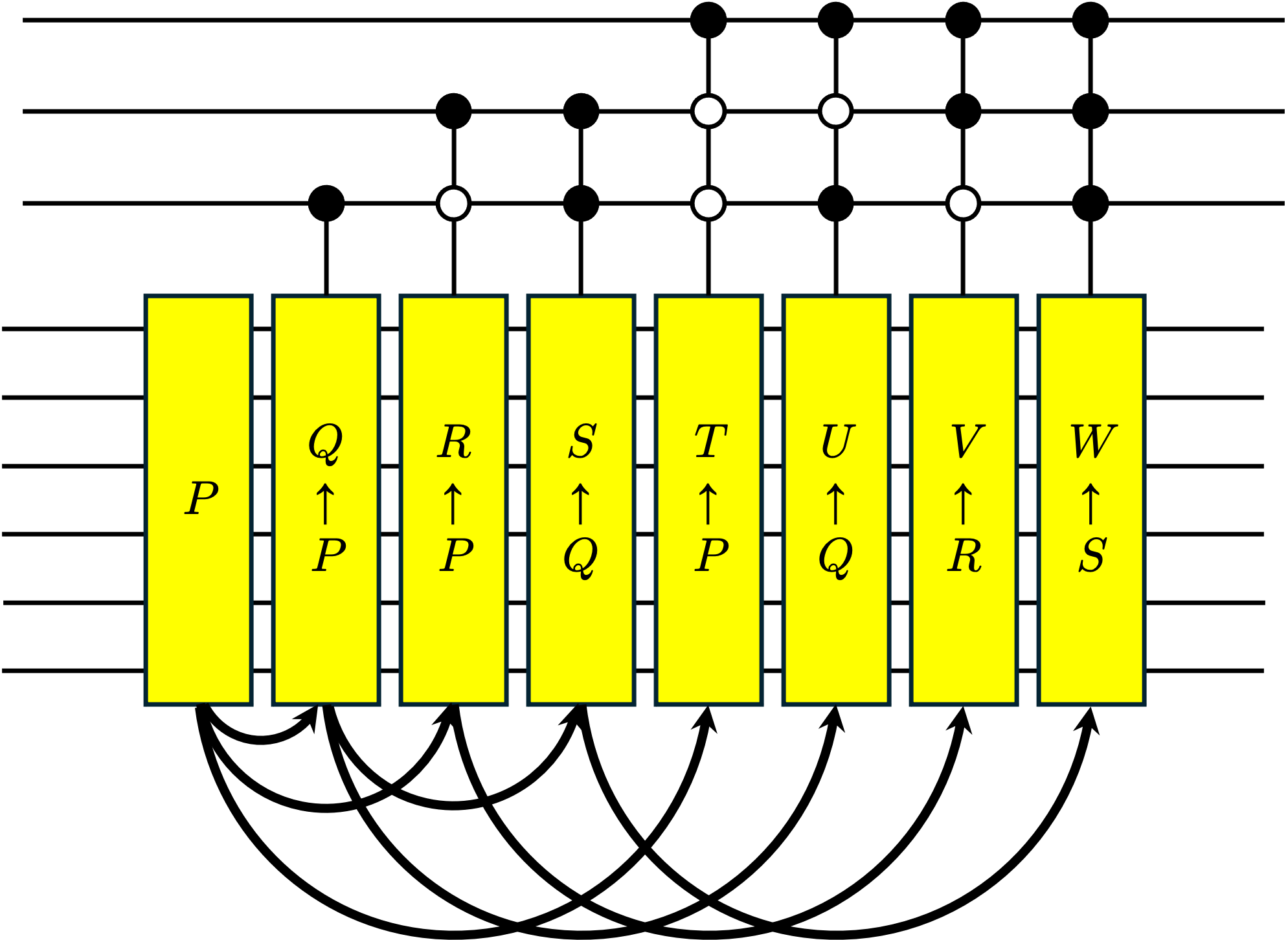}
\caption{Controlled gate network approach for producing a linear combination of eight unitary operators using three ancilla qubits.}
\label{fig:LCU_controlled}
\end{figure}

If we use a controlled gate network to produce the linear combination of unitary operators, then we can omit the conditional control for any leading $0$'s appearing in our control-state binary digits.  This is illustrated in Fig.~\ref{fig:LCU_controlled} for $N = 8$ unitary operators and $n = 3$ ancilla qubits.  It is straightforward to show that the number of removed controlling operators is $2^n-1=N-1$.  Since these controlling operations are applied to each of the simpler unitary operators in Eq.~\eqref{eq:CNOTreduction}, the reduction in the number of CNOT gates can be very significant.  Suppose that at some step $k$, none of the gates $P_k, Q_k, R_k, S_k$ are proportional to the identity, and we satisfy the conditions
\begin{align}
    & \nu(C(P^\dagger_k Q_k)) \le \nu(C(Q_k)), \;
    \nu(C(C(P^\dagger_k R_k))) \le \nu(C(C(R_k))), \nonumber \\
  & \nu(C(C(Q^\dagger_k S_k))) \le \nu(C(C(S_k))), \;  
      \nu(C(C(C(P^\dagger_k T_k)))) \le \nu(C(C(C(T_k)))), 
   \nonumber \\ 
 &  \nu(C(C(C(Q^\dagger_k U_k)))) \le \nu(C(C(C(U_k)))), \;
      \nu(C(C(C(R^\dagger_k V_k)))) \le \nu(C(C(C(V_k)))), 
   \nonumber \\ 
   &   \nu(C(C(C(S^\dagger_k W_k)))) \le \nu(C(C(C(W_k)))).
\end{align}
These conditions are analogous to those presented in Eq.~\eqref{eq:CNOTreduction} and are satisfied when $P_k, \cdots, W_k$ touch the same qubits and don't have simplifying structures that make the controlled operations for $P_k, \cdots, W_k$ simpler than those of the bilinear combinations $P^\dagger_k Q_k, \cdots S^\dagger_k W_k$.  When these conditions are satisfied, it is straightforward to show that the controlled gate network method reduces the number of CNOT gates by at least $2^3-1 = 7$.  When the analogous conditions hold for a system with $N=2^n$ unitary operators and $n$ ancilla qubits, the reduction in the number of CNOT gates at step $k$ will be at least $2^n-1 = N-1$.

\section{Applications to variational subspace calculations}
In the quantum approximate optimization algorithm (QAOA) \cite{farhi2014quantum,farhi2016quantum,zhou2020quantum}, one approximates the ground state of a Hamiltonian $H$ using a variational ansatz of the form, 
\begin{equation}
\ket{\vec{\epsilon},\vec{\lambda}} = \exp(-i\epsilon_N H_I)\exp(-i\lambda_N H) \cdots 
    \exp(-i\epsilon_1 H_I)\exp(-i\lambda_1 H)\ket{\psi_I},
\end{equation}
where $\ket{\psi_I}$ is the ground state of some trivial initial Hamiltonian, $H_I$.  The parameters $\vec{\epsilon}$ and $\vec{\lambda}$ are then optimized to give the lowest expectation value $\braket{\vec{\epsilon},\vec{\lambda}|H|\vec{\epsilon},\vec{\lambda}}$ \footnote{These parameters are typically labeled $\vec{\beta}$ and $\vec{\gamma}$ in the QAOA literature; however, these symbols have already been used elsewhere in the text so they have been relabeled here for clarity.}.  However, this single-vector ansatz for the ground state may not be sufficient.  The variational approximation can be improved if we allow for linear combinations of variationally-optimized QAOA vectors $\ket{\vec{\epsilon}^{(i)},\vec{\lambda}^{(i)}}$.  Variational optimization using a subspace rather than a single vector allows for the mixing of configurations that explicitly break symmetries of the Hamiltonian such as rotational invariance, particle number conservation, or internal flavor symmetries. The underlying symmetry is restored by allowing for linear combinations of the individual configurations.  This strategy of explicit symmetry breaking and symmetry restoration is widely used in nuclear physics 
\cite{Satula:2009sx,Yao:2019rck,Sheikh:2019qdz,Frosini:2021sxj,RuizGuzman:2021qyj,Lacroix:2022vmg}.  In order to perform a variational subspace calculation, one needs to be able to compute inner products, $\braket{\vec{\epsilon}^{(i)},\vec{\lambda}^{(i)}|\vec{\epsilon}^{(j)},\vec{\lambda}^{(j)}}$ and Hamiltonian matrix elements $\braket{\vec{\epsilon}^{(i)},\vec{\lambda}^{(i)}|H|\vec{\epsilon}^{(j)},\vec{\lambda}^{(j)}}$ between pairs of QAOA states.  One can then solve the generalized eigenvalue problem using classical post-processing to find the ground state of $H$ within the projected subspace \cite{Frame:2017fah,Epperly:2021ugt,2022arXiv220910571F,Mejuto-Zaera:2023ier,Sarkar:2023qjn,Duguet:2023wuh}.

To illustrate with a simple example, we consider a two-qubit system where the Hamiltonian $H$ corresponds to a Heisenberg model, 
\begin{equation}
    H = a X_1 X_2 + b Y_1 Y_2 + c Z_1 Z_2.
\end{equation}
If we take the initial Hamiltonian to be $H_I = -\sum_i X_i$, then each layer of the QAOA product will produce a product of exponentials of the form,
\begin{equation}
    N(\alpha,\beta,\gamma,\delta) =\exp(i\delta X_1)\exp(i\delta X_2) \exp[i(\alpha X_1 X_2 + \beta Y_1 Y_2 + \gamma Z_1 Z_2)],
\end{equation}
with real parameters $\alpha$, $\beta$, $\gamma$, $\delta$.
In order to perform the variational subspace calculations using such QAOA states, we need to construct superpositions of $\ket{A}\equiv N(\alpha,\beta,\gamma,\delta)\ket{\psi_I}$ and $\ket{B}\equiv N(\alpha+\Delta \alpha,\beta+\Delta \beta,\gamma+\Delta \gamma,\delta+\Delta\delta)\ket{\psi_I}$, where $\ket{\psi_I}$ is the initial state of the system.  We note that this two-qubit example is only intended as a pedagogical example to illustrate how variational subspace calculations are performed on larger systems with more than two qubits.  Let us first show how this is done efficiently using a controlled gate network.

We can implement $N(\alpha,\beta,\gamma,\delta)$ and $N(\alpha+\Delta \alpha,\beta+\Delta \beta,\gamma+\Delta \gamma,\delta+\Delta\delta)$ using a controlled gate network as shown in Fig.~\ref{fig:variational}.  We are using the implementation of $N(\alpha,\beta,\gamma,\delta)$ described in Ref.~\cite{Smith2019}.  The transformation gates are highlighted in yellow.  When the transformation gates are not included, the circuit gives $N(\alpha,\beta,\gamma,\delta)$.  When the transformation gates are included, the circuit equals $N(\alpha+\Delta \alpha,\beta+\Delta \beta,\gamma+\Delta \gamma,\delta+\Delta\delta)$. Using an ancillary qubit to control all the transformation gates, we can construct an arbitrary superposition of $\ket{A}\equiv N(\alpha,\beta,\gamma,\delta)\ket{\psi_I}$ and $\ket{B}\equiv N(\alpha+\Delta \alpha,\beta+\Delta \beta,\gamma+\Delta \gamma,\delta+\Delta\delta)\ket{\psi_I}$, where $\ket{\psi_I}$ is the initial state of the system.

\begin{figure}[ht]
\begin{adjustbox}{width=13cm}
\begin{quantikz}[transparent,slice style=blue]
\lstick{} & \gate{R_{x/y}(\theta)}\gategroup[1,steps=1,style={dashed,fill=cyan!60, inner xsep=2pt},background,label style={label position=below,anchor=north,yshift=-0.2cm}]{{\sc }}  & \qw & \qw & \ctrl{1} & \ctrl{2} & \qw & \qw & \ctrl{2} & \qw  & \qw &\qw & \ctrl{2}& \ctrl{1} & \gate{H} & \meter{}  \\
\qw & \qw    &  \targ{} & \gate{R_z(\pi/2 -2\gamma)} & \gate{R_z(-2\Delta\gamma)}\gategroup[1,steps=1,style={fill=yellow,inner sep=2pt},background] \qw & \qw & \ctrl{1} & \qw & \qw & \targ{} & \gate{R_z(\pi/2)} &  \gate{R_x(-2\delta)}  &\qw &\gate{R_x(-2\Delta\delta)}\gategroup[1,steps=1,style={fill=yellow,inner sep=2pt},background]  \qw & \qw &\qw\\
\qw & \gate{R_z(-\pi/2)} & \ctrl{-1} &  \gate{R_y(2\alpha-\pi/2)} & \qw & \gate{R_y(2\Delta\alpha)}\gategroup[1,steps=1,style={fill=yellow,inner sep=2pt},background]  \qw & \targ{} & \gate{R_y(\pi/2-2\beta)} & \gate{R_y(-2\Delta\beta)}\gategroup[1,steps=1,style={fill=yellow,inner sep=2pt},background]  \qw & \ctrl{-1} & \gate{R_x(-2\delta)} & \qw &\gate{R_x(-2\Delta\delta)}\gategroup[1,steps=1,style={fill=yellow,inner sep=2pt},background]  \qw & \qw & \qw & \qw
\end{quantikz}
\end{adjustbox}
\caption{Controlled gate network for our variational subspace calculation example. The transformation gates are highlighted in yellow.  When the transformation gates are not included, the circuit gives $N(\alpha,\beta,\gamma,\delta)$.  When the transformation gates are included, the circuit equals $N(\alpha+\Delta \alpha,\beta+\Delta \beta,\gamma+\Delta \gamma,\delta+\Delta\delta)$.  By using one ancilla qubit to control all of the transformation gates, we can construct an arbitrary superposition of $N(\alpha,\beta,\gamma,\delta)$ and $N(\alpha+\Delta \alpha,\beta+\Delta \beta,\gamma+\Delta \gamma,\delta + \Delta\delta)$.}
\label{fig:variational}
\end{figure}

By aligning $\hat n$ with $y$ and $x$ axes in the $R_{\hat n}(\theta)$ gate and performing a measurement on the ancilla qubit, we can deduce the real and imaginary parts of the inner product $\braket{A|B}$ from the measurement probabilities of the $\ket{0}$ and $\ket{1}$ states.  We can apply the same technique to calculate the matrix elements for the Hamiltonian, $\langle A|H|B\rangle$, and their complex conjugates.  This is done by decomposing $H$ into a sum of unitary operators such as strings of Paul matrices, $H=\sum_k U_k$.  We then replace $\ket{B}$ in the description above with $U_k\ket{B}$ for each $k$.  If we decompose and transpile the circuit in Fig.~\ref{fig:variational} into native gates for the IBM system, we get the circuit shown in Fig.~\ref{fig:variational_transpiled}.  A total of 13 CNOT gates and 21 single qubit gates are required.

\begin{figure}[ht]
\begin{adjustbox}{width=13cm} % Adjust the scale factor as needed
\begin{quantikz}[transparent, slice style=blue, every arrow/.append style={-Triangle, line width=0.7mm}]
\lstick{} & \gate{U_3(\theta,-\frac{\pi}{2}/0,\frac{\pi}{2}/0)}\gategroup[1,steps=1,style={dashed,fill=cyan!60, inner xsep=2pt},background,label style={label position=below,anchor=north,yshift=-0.2cm}]{{\sc }}  & \qw & \qw & \qw\gategroup[2,steps=4,style={dashed,rounded corners,fill=yellow, inner xsep=2pt},background,label style={label position=below,anchor=north,yshift=-0.2cm}]{{\sc }}  & \ctrl{1} & \qw & \ctrl{1} & \qw \gategroup[3,steps=4,style={dashed,rounded corners,fill=yellow, inner xsep=2pt},background,label style={label position=below,anchor=north,yshift=-0.2cm}]{{\sc }}  & \ctrl{2} & \qw &  \ctrl{2} & \qw & \qw & \qw \gategroup[3,steps=4,style={dashed,rounded corners,fill=yellow, inner xsep=2pt},background,label style={label position=below,anchor=north,yshift=-0.2cm}]{{\sc }}  & \ctrl{2}&  \qw   & \ctrl{2} &  \arrow[r]\qw & \\
\qw & \qw    &  \targ{} & \gate{U_1(\pi/2 -2\gamma)} & \gate{U_1(-\Delta\gamma)}& \targ{} &\gate{U_1(\Delta\gamma)}& \targ{} &\qw &\qw &\qw & \qw &\ctrl{1}  & \qw &   \qw & \qw & \qw &\qw &   \arrow[r]\qw&  \\
\qw &\gate{U_1(-\pi/2)} & \ctrl{-1} &  \gate{U_3(2\alpha-\pi/2,0,0)} & \qw &\qw &\qw &\qw& \gate{U_3(\Delta\alpha,0,0)}& \targ{}&\gate{U_3(-\Delta\alpha,0,0)} & \targ{} & \targ{} & \gate{U_3(\pi/2-2\beta,0,0)} & \gate{U_3(-\Delta\beta,0,0)}& \targ{}& \gate{U_3(\Delta\beta,0,0)}& \targ{}&  \arrow[r]\qw&
\end{quantikz}
\end{adjustbox}

\vspace{0.5cm}

\begin{adjustbox}{width=13cm}
\begin{quantikz}[transparent, slice style=blue, every arrow/.append style={-Triangle, line width=0.7mm}]
\qw & \arrow[l]&  \qw & \qw& \qw & \qw & \qw\gategroup[3,steps=5,style={dashed,rounded corners,fill=yellow, inner xsep=2pt},background,label style={label position=below,anchor=north,yshift=-0.2cm}]{{\sc }}  &  \ctrl{2}& \qw  &  \ctrl{2}& \qw& \qw& \qw\gategroup[2,steps=5,style={dashed,rounded corners,fill=yellow, inner xsep=2pt},background,label style={label position=below,anchor=north,yshift=-0.2cm}]{{\sc }}  &\ctrl{1} & \qw & \ctrl{1} &\qw&\gate{U_2(0,\pi)} & \meter{}  \\
\qw & \arrow[l]& \qw & \targ{} & \gate{U_1(\pi/2)} &  \gate{U_3(-2\delta,-\frac{\pi}{2},\frac{\pi}{2})}  &\qw & \qw & \qw &\qw&\qw & \qw&\gate{U_1(\pi/2)} & \targ{}&\gate{U_3(\Delta\delta,0,0)} & \targ{}&\gate{U_3(-\Delta\delta,-\frac{\pi}{2},0)}& \qw & \qw \\
\qw & \arrow[l]& \qw & \ctrl{-1} & \gate{U_3(-2\delta,-\frac{\pi}{2},\frac{\pi}{2})} & \qw &\gate{U_1(\pi/2)} & \targ{}&\gate{U_3(\Delta\delta,0,0)} & \targ{}&\gate{U_3(-\Delta\delta,-\frac{\pi}{2},0)}& \qw & \qw & \qw& \qw & \qw& \qw & \qw & \qw
\end{quantikz}
\end{adjustbox}
\caption{This is the fully decomposed and transpiled version for the circuit in Fig.~\ref{fig:variational} using gates native to the IBM system.  A total of 13 CNOT gates and 21 single qubit gates are required.} 
\label{fig:variational_transpiled}
\end{figure}

We can now compare with the standard approach for calculating $\braket{A|B}$ using a Hadamard test \cite{2022arXiv220910571F}.  The corresponding circuit is shown in Fig.~\ref{fig:standard}.  
If we decompose and transpile the circuit in Fig.~\ref{fig:standard} into native gates for the IBM quantum devices, we get the circuit shown in Fig.~\ref{fig:standard_transpiled}.  A total of 64 CNOT gates and 82 single qubit gates are required.  We conclude that the controlled gate network approach is substantially more efficient than the standard approach.  The reduction in the number of gates is a factor of about $5$ for the CNOT gates and a factor of about $4$ for the single qubit gates. 

\begin{figure}
\centering
\begin{adjustbox}{width=10cm}
\begin{quantikz}[transparent,slice style=blue]
 & \gate{H}      & \ctrl{1}
% \gategroup[2,steps=3,style={dashed, rounded corners,fill=cyan!20, inner xsep=2pt},background,label style={label position=below,anchor= north,yshift=-0.2cm}]{{Base system}}    
& \qw & \ctrl{1} & \gate{R_{x/y}(\theta)}\gategroup[1,steps=1,style={dashed,fill=cyan!60, inner xsep=2pt},background,label style={label position=below,anchor=north,yshift=-0.2cm}]{{\sc }}   & \meter{} \\
 & \qw & \gate{N(\alpha,\beta,\gamma,\delta)}  & \qw & \gate{N^\dagger(\alpha+\Delta\alpha,\beta+\Delta\beta,\gamma+\Delta\gamma,\delta+\Delta\delta)}  & \qw & \qw
\end{quantikz}
\end{adjustbox}
\caption{Standard Hadamard test circuit for computing the real and imaginary parts of the inner product $\braket{A|B}$.}
\label{fig:standard}
\end{figure}

\begin{figure}
\begin{adjustbox}{width=13cm}
\begin{quantikz}[transparent, slice style=blue, every arrow/.append style={-Triangle, line width=0.7mm}]
\qw& \gate{U_2(0,\pi)} & \ctrl{2} & \qw & \ctrl{2} & \qw& \qw& \ctrl{1} & \qw & \qw& \qw& \ctrl{1} &\qw &\ctrl{2} &\gate{U_1(\pi/4)}&\ctrl{2} &\ctrl{1} &\qw &\ctrl{1} &\ctrl{2} &\qw &\ctrl{2}&  \arrow[r]\qw&
\\
\qw&\qw & \qw & \gate{U_2(0,\pi)} & \qw  &\targ{} & \gate{U_1(-\pi/4)}&\targ{} & \gate{U_1(\pi/4)} &\targ{}& \gate{U_1(-\pi/4)}&\targ{} & \gate{U_2(0,-3\pi/4)} &\qw &\gate{U_1(\pi/4-\gamma)} &\qw &\targ{} &\gate{U_1(\gamma-\pi/4)} & \targ{} &\qw&\qw &\qw&  \arrow[r]\qw&
\\
\qw&\gate{U_1(-\pi/4)} & \targ{} & \gate{U_1(\pi/4)} & \targ{} & \ctrl{-1}& \qw& \qw& \qw& \ctrl{-1} & \gate{U_1(\pi/4)}& \qw & \qw & \targ{} &\gate{U_1(-\pi/4)} &\targ{} &\qw  &\gate{U_3(\alpha-\pi/4,0,0)} &\qw &\targ{} &\gate{U_3(\pi/4-\alpha,0,0)} &\targ{}&  \arrow[r]\qw&
\end{quantikz}
\end{adjustbox}

\vspace{0.5cm}

\begin{adjustbox}{width=13cm}
\begin{quantikz}[transparent, slice style=blue, every arrow/.append style={-Triangle, line width=0.7mm}]
\qw & \arrow[l]& \qw & \qw& \qw& \ctrl{2}& \qw& \qw& \qw& \ctrl{2} &\ctrl{1} &\qw &\gate{U_1(\pi/4)} &\ctrl{1} &\ctrl{2} &\qw &\ctrl{2}
&\qw&\qw &\ctrl{1} &\qw &\qw &\qw &\arrow[r]\qw&
\\
\qw & \arrow[l]& \qw &\ctrl{1} &\qw &\qw&\qw &\ctrl{1}  &\gate{U_1(\pi/4)}&\qw &\targ{} &\qw &\gate{U_1(-\pi/4)} &\targ{} &\qw &\gate{U_2(0,\pi)} &\qw &\targ{} &\gate{U_1(-\pi/4)} &\targ{} &\gate{U_1(\pi/4)} &\targ{}&\gate{U_1(-\pi/4)}  &\arrow[r]\qw&
\\
\qw & \arrow[l]& \gate{U_2(0,\pi)} & \targ{} &\gate{U_1(-\pi/4)} & \targ{} &\gate{U_1(\pi/4)} & \targ{} &\gate{U_1(-\pi/4)} &\targ{} &\gate{U_2(0,-3\pi/4)} &\qw &\gate{U_3(\pi/4-\beta, 0,0)}&\qw &\targ{} &\gate{U_3(\beta-\pi/4, 0,0)} &\targ{} &\ctrl{-1} &\qw &\qw &\qw &\ctrl{-1} &\gate{U_1(\pi/4)} &\arrow[r]\qw&
\end{quantikz}
\end{adjustbox}

\vspace{0.5cm}

\begin{adjustbox}{width=13cm}
\begin{quantikz}[transparent, slice style=blue, every arrow/.append style={-Triangle, line width=0.7mm}]
\qw & \arrow[l]& \ctrl{1} &\qw & \ctrl{2} &\gate{U_1(\frac{\pi}{4})} & \ctrl{2}& \ctrl{1} &\qw & \ctrl{1} &\ctrl{2} &\qw & \ctrl{2} & \ctrl{1} &\qw & \ctrl{1} &\qw &\qw & \ctrl{1} &\qw & \ctrl{1} &\qw &\arrow[r]\qw&
\\
\qw & \arrow[l]& \targ{} &\gate{U_2(0,-\frac{3\pi}{4})} & \qw & \gate{U_1(\frac{\pi}{4})} & \qw & \targ{} & \gate{U_1(-\frac{\pi}{4})} & \targ{} &\qw &\gate{U_1(\frac{\pi}{2})}  &\qw &\targ{} &\gate{U_3(\delta,0,0)} &\targ{} &\gate{U_3(-\delta, -\frac{\pi}{2},0)} & \gate{U_1(\frac{\pi}{2})} &\targ{} &\gate{U_3(-\Delta\delta-\delta,0,0)} &\targ{} &\gate{U_3(\Delta\delta+\delta,-\frac{\pi}{2},0)} &\arrow[r]\qw&
\\
\qw & \arrow[l]& \qw & \qw & \targ{}  &\gate{U_1(-\frac{\pi}{4})} & \targ{}& \qw &\gate{U_1(\frac{\pi}{2})} & \qw & \targ{} &\gate{U_3(\delta,0,0)} & \targ{} &\gate{U_3(-\delta, -\frac{\pi}{2},0)} &\qw &\qw &\qw &\qw &\qw &\qw  &\qw  &\qw &\arrow[r]\qw&
\end{quantikz}
\end{adjustbox}

\vspace{0.5cm}

\begin{adjustbox}{width=13cm}
\begin{quantikz}[transparent, slice style=blue, every arrow/.append style={-Triangle, line width=0.7mm}]
\qw & \arrow[l]& \qw &\ctrl{2} & \ctrl{1}  &\ctrl{2}  & \qw & \ctrl{1} & \qw & \qw & \qw & \ctrl{1} & \qw & \qw & \qw & \ctrl{1} & \ctrl{2} & \gate{U_1(\frac{\pi}{4})} & \ctrl{2} & \qw & \ctrl{2} & \qw &\arrow[r]\qw&
\\
\qw & \arrow[l]& \gate{U_1(-\frac{\pi}{4})} & \qw & \targ{} & \qw  & \gate{U_1(\frac{\pi}{4})} & \targ{}  & \gate{U_2(0, \pi)} & \targ{}  & \gate{U_1(-\frac{\pi}{4})} & \targ{} & \gate{U_1(\frac{\pi}{4})} & \targ{} & \gate{U_1(-\frac{\pi}{4})} & \targ{} &\qw & \gate{U_2(0,-\frac{3\pi}{4})} &\qw
& \qw & \qw & \qw  &\arrow[r]\qw&
\\
\qw & \arrow[l]& \gate{U_1(\frac{\pi}{2})} & \targ{} & \gate{U_3(-\Delta\delta-\delta,0,0)}  & \targ{} & \gate{U_3(\Delta\delta+\delta,-\frac{\pi}{2},0)} &\qw &\qw & \ctrl{-1} & \qw & \qw & \qw & \ctrl{-1} & \gate{U_1(\frac{\pi}{4})} & \qw & \targ{} & \gate{U_1(-\frac{\pi}{4})} & \targ{} &\gate{U_3(\Delta\beta+\beta-\frac{\pi}{4},0,0)} & \targ{} &\gate{U_3(-\Delta\beta-\beta+\frac{\pi}{4},0,0)} &\arrow[r]\qw& 
\end{quantikz}
\end{adjustbox}

\vspace{0.5cm}

\begin{adjustbox}{width=13cm}
\begin{quantikz}[transparent, slice style=blue, every arrow/.append style={-Triangle, line width=0.7mm}]
\qw & \arrow[l]& \ctrl{2} &\qw &\qw & \qw & \ctrl{2} & \qw & \qw & \qw & \ctrl{2} &\ctrl{1} & \gate{U_1(\frac{\pi}{4})} &\ctrl{1} &\ctrl{2} &\qw &\ctrl{2} &\ctrl{1} & \qw &\ctrl{1} & \qw & \qw  &\arrow[r]\qw& 
\\
\qw & \arrow[l]& \qw & \qw &\ctrl{1} & \qw & \qw & \qw &\ctrl{1} &\gate{U_1(\frac{\pi}{4})} & \qw &\targ{} & \gate{U_1(-\frac{\pi}{4})}&\targ{} &\qw &\gate{U_1(\Delta\gamma+\gamma-\frac{\pi}{4})} &\qw &\targ{} &\gate{U_1(-\Delta\gamma-\gamma+\frac{\pi}{4})} &\targ{} &\gate{U_2(0,\pi)} &\targ{}  &\arrow[r]\qw& 
\\
\qw & \arrow[l]& \targ{} &\gate{U_2(0,\pi)} & \targ{} &\gate{U_1(-\frac{\pi}{4})}  & \targ{} &\gate{U_1(\frac{\pi}{4})} & \targ{} &\gate{U_1(-\frac{\pi}{4})} & \targ{} &\gate{U_2(0,-\frac{3\pi}{4})} &\gate{U_3(-\Delta\alpha-\alpha+\frac{\pi}{4},0,0)} &\qw &\targ{} &\gate{U_3(\Delta\alpha+\alpha-\frac{\pi}{4},0,0)} &\targ{} &\qw &\qw &\qw &\qw &\ctrl{-1}  &\arrow[r]\qw& 
\end{quantikz}
\end{adjustbox}

\vspace{0.5cm}

\begin{adjustbox}{width=13cm}
\begin{quantikz}[transparent, slice style=blue, every arrow/.append style={-Triangle, line width=0.7mm}]
\qw & \arrow[l]& \qw &\ctrl{1} & \qw & \qw & \qw  & \ctrl{1} & \qw & \ctrl{2} &\gate{U_1(\frac{\pi}{4})} & \ctrl{2} & \qw & \ctrl{2} & \qw & \ctrl{2} & \qw  & \qw & \gate{U_3(\theta,-\frac{\pi}{2}/0,\frac{\pi}{2}/0)}\gategroup[1,steps=1,style={dashed,fill=cyan!60, inner xsep=2pt},background,label style={label position=below,anchor=north,yshift=-0.2cm}]{{\sc }}  &\meter{}
\\
\qw & \arrow[l]& \gate{U_1(-\frac{\pi}{4})} &\targ{} & \gate{U_1(\frac{\pi}{4})} &\targ{} & \gate{U_1(-\frac{\pi}{4})} &\targ{} &\gate{U_2(0,-\frac{3\pi}{4})} & \qw & \qw& \qw & \qw & \qw & \qw & \qw & \qw & \qw  & \qw & \qw
\\
\qw & \arrow[l]& \qw & \qw & \qw & \ctrl{-1} &\gate{U_1(\frac{\pi}{4})}& \qw & \qw &\targ{}&\gate{U_1(-\frac{\pi}{4})} &\targ{} &\gate{U_1(\frac{\pi}{4})} &\targ{} &\gate{U_1(-\frac{\pi}{4})} &\targ{}  & \qw & \qw  & \qw & \qw
\end{quantikz}
\end{adjustbox}
\caption{The fully decomposed and transpiled version for the circuit in Fig.~\ref{fig:standard} using gates native to the IBM quantum devices.  A total of 64 CNOT gates and 82 single qubit gates are required.}
\label{fig:standard_transpiled}
\end{figure}

\section{Applications to rodeo algorithm calculations}
\subsection{Rodeo algorithm}
There are several well-known quantum computing algorithms that measure the eigenvalues of a Hermitian operator $H$ using time evolution controlled by an auxiliary quantum register. An important example is the class of phase estimation algorithms, which includes standard phase estimation \cite{Abrams:1999,Cleve:1998,Siwach:2021krs} and iterative quantum phase estimation \cite{Kitaev:1995qy,Svore:2013}. The rodeo algorithm is another recently introduced method that uses controlled time evolution \cite{Choi:2020pdg}. It is an iterative algorithm that uses destructive interference to suppress eigenvectors with eigenvalues different from a desired ``target" eigenvalue. The rodeo algorithm was shown to have exponentially improved scaling in terms of the precision $\Delta$ for preparing eigenstates \cite{Choi:2020pdg}, compared to quantum phase estimation and adiabatic evolution \cite{Farhi:2000a,Wiebe:2011a}. 

The first implementation of the rodeo algorithm was performed in Ref.~\cite{Qian:2021wya} for a single qubit Hamiltonian. However, the use of controlled time evolution makes the implementation for a multi-qubit system significantly more difficult, especially on existing noisy quantum devices.  In order to reduce the number of two qubit operations, in this work we use controlled reversal gates that allow one to flip the direction of time evolution using an auxiliary qubit.

Let us consider the two-qubit Hamiltonian 
\begin{align}\label{eq:system}
    H_\mathrm{obj} = c_1 X_1 Z_2 + c_2 Z_1  X_2
\end{align}
where \(c_1, c_2 \in \mathbb{R}\) and \(X_i\) and \(Z_i\) are the Pauli operators on qubit $i$. Following the notation of Ref.~\cite{Choi:2020pdg,Qian:2021wya}, we call $H_\mathrm{obj}$ the ``object Hamiltonian". We take the coefficients to be $c_1 = 2.5$ and $c_2 = 1.5$, and the Hamiltonian has four eigenvalues \(\pm(c_1 \pm c_2)\). The quantum circuit that implements the time evolution of \(H_\mathrm{obj}\) is shown in \cref{fig:system}.

Each cycle of the rodeo algorithm uses a controlled time evolution with a random time $t$ sampled from a normal distribution with zero mean and standard deviation \(\sigma\) set by the user. For any given rodeo cycle, the measurement of \(0\) for the ancilla corresponds to a successful measurement. Let $\ket{\psi_I}$ be the initial state and let $\ket{E_k}$ denote the energy eigenstates of $H_{\rm obj}$ with energy $E_k$. If we apply \(n\) cycles of the rodeo algorithm and average over many random trials, the probability of success for all $n$ cycles at target energy $E$ (a free parameter) is 
\begin{equation}
    P_n(E) = \sum_k  \frac{\left[1+e^{-(E-E_k)^2\sigma^2/2}\right]^n \left| \braket{E_k|\psi_I} \right|^2 }{2^n} .
    \label{eq:probability1}
\end{equation}
We see that there are peaks centered at each energy eigenvalue $E_k$ where the initial state has nonzero overlap with $\ket{E_k}$. We also see that there is a constant background value of \(\frac{1}{2^n}\) for values of $E$ that are significantly further away than $\sigma^{-1}$ from all energy eigenvalues. Thus, repeated cycles will indicate peaks whenever $E$ is chosen within \(O(\sigma^{-1})\) of some \(E_k\) such that \(\left| \braket{E_k|\psi_I} \right|^2\) is not vanishingly small.

When implemented using the IBM Qiskit transpiler \cite{Qiskit} under the assumption of full connectivity between all qubits, directly controlling the time evolution of the circuit in \cref{fig:system} with an ancilla qubit requires a total of 20 CNOT gates. For the Quantinuum H1-2 device at the time of this experiment, the two-qubit entangling gate was the $R_{ZZ}$ gate with angle $\pi/2$, and 20 $R_{ZZ}$ gates are needed for the analogous circuit. With only partial connectivity, the number of two-qubit entangling gates grows even larger. For three cycles of the rodeo algorithm, we therefore need 60 two-qubit entangling gates. For five cycles of the rodeo algorithm, we need 100 two-qubit entangling gates. For systems with more than two qubits, the problem is even more severe since the number of two-qubit gates will also scale with the number of Trotter-Suzuki time steps. 

\begin{figure}
\centering
\begin{adjustbox}{width=6cm}
\begin{quantikz}[transparent,slice style=blue]
\lstick{$\ket{0}$} & \qw      & \ctrl{1}
% \gategroup[2,steps=3,style={dashed, rounded corners,fill=cyan!20, inner xsep=2pt},background,label style={label position=below,anchor= north,yshift=-0.2cm}]{{Base system}}    
& \gate{R_x(2 c_1 t)} & \ctrl{1} & \qw      & \qw \\
\lstick{$\ket{0}$} & \gate{H} & \targ{}  & \gate{R_z(2 c_2 t)} & \targ{}  & \gate{H} & \qw
\end{quantikz}
\end{adjustbox}
\caption{Quantum circuit implementing the two-qubit unitary \(\exp\{-i H_\mathrm{obj} t\}\), where \(H_\mathrm{obj}\) is given by~\cref{eq:system}.}
\label{fig:system}
\end{figure}

\subsection{Controlled reversal gates}
Controlled reversal gates can reduce these circuit complexities. We define a reversal gate $R$ to be a product of single qubit gates $R = G_1 G_2 \cdots G_N$ that anticommutes with some subset of terms in the Hamiltonian, and write $H_R$ to be the part of the Hamiltonian anticommuting with $R$. If the Hamiltonian is written as a sum of products of Pauli gates, then it is straightforward to partition the terms in the Hamiltonian so that each partition has some corresponding reversal gate. A controlled reversal gate $C(R)$ is the controlled version of $R$, i.e., $C(R) = C(G_1)C(G_2) \cdots C(G_N)$. Since $R H_R R^\dagger = -H_R$, and since $-H$ generates a backward evolution of $H$, $C(R)$ allows one to toggle the flow of time forward and backward, depending on the state of the auxiliary register. 

The relative phase between the states entangled with the ancilla states $\ket{0}$ and $\ket{1}$ is the physically relevant quantity, while any global phase is irrelevant.  When using controlled reversal gates, it is desirable that the approximate time evolution operator $\tilde{U}(t)$ is time-reversal symmetric, in the sense of $\tilde{U}(-t) = \tilde{U}(t)^\dagger$. This ensures that $\tilde{U}(t)$ and $\tilde{U}(-t)$ have the same stationary states. For our two-qubit Hamiltonian example, the evolution is exact and thus this symmetry is satisfied. For Hamiltonian simulation using the Trotter approximation, the symmetry is satisfied if the single Trotter step $\mathcal{P}(t)$ satisfies $\mathcal{P}(-dt)=\mathcal{P}(dt)^{-1}$. This is automatically satisfied by the recursive Trotter-Suzuki formulas, particularly at second order \cite{Suzuki:1993,Hatano:2005}. 

The controlled reversal gates implement forward or backward time evolution depending on the ancilla state.  Therefore, the relative phase between the two states evolves at twice the rate as the standard controlled-$\tilde{U}$ setup, and we can reduce the number of Trotter steps by a factor of two. This feature alone provides a two-fold performance advantage.  For this reason, the computational advantage in using controlled reversal gates with the rodeo algorithm is at least a factor of two, independent of the details of the Hamiltonian.  For the Hamiltonian that we consider here, we can find a reversal gate that commutes with all of the terms in the Hamiltonian, resulting in an even greater computational advantage.  For the general case, however, we will need to break up the Hamiltonian into pieces and apply the appropriate reversal gate for each piece.  If we are using the Trotter approximation to implement time evolution, then this breakup of the Hamiltonian is already done, and the application of controlled reversal gates for the time evolution is straightforward.  

\subsection{Identifying energy eigenvalues}

We now use controlled reversal gates and the rodeo algorithm to compute the energy spectrum of $H_\mathrm{obj}$ from Eq.~\eqref{eq:system}. We note that the single-qubit Pauli \(Y_1\) anticommutes with $H_\mathrm{obj}$ and thus choose $C_{Y_1}$ as our controlled reversal gate ($C_{Y_2}$ is equally valid). One cycle of the rodeo algorithm for our system is shown in \cref{fig:1CycleCirc}. It consists of a pair of controlled reversal gates controlled by a single ancilla qubit followed by mid-circuit measurement. By implementing controlled reversal gates, we are able to reduce the number of CNOT gates for each rodeo cycle from 20 to 4.

\begin{figure}
\centering
\begin{adjustbox}{width=10cm}

\begin{quantikz}[transparent,slice style=blue]
\lstick{$\ket{0}$} & \gate{H} & \octrl{1}
\gategroup[2,steps=1,style={dashed, rounded corners,fill=cyan!20, inner xsep=2pt},background,label style={label position=above,anchor= west,xshift=0.75cm}]{{Controlled reversal gates}} 
                                          & \qw      & \qw                 & \qw      & \octrl{1}  
\gategroup[2,steps=1,style={dashed, rounded corners,fill=cyan!20, inner xsep=2pt},background]{} &
\gate{P(2E t)} & \gate{H} & \meter{} & \qw \\
\lstick{$\ket{0}$} & \qw      & \gate{Y}  & \ctrl{1} & \gate{R_x(2 c_1 t)} & \ctrl{1} & \gate{Y}  &  \qw             & \qw      & \qw      & \qw \\
\lstick{$\ket{0}$} & \qw      & \gate{H}  & \targ{}  & \gate{R_z(2 c_2 t)} & \targ{}  & \gate{H}  &  \qw             & \qw      & \qw      & \qw 
\end{quantikz}
\end{adjustbox}
\caption{Circuit diagram for one cycle of the rodeo algorithm on the two-qubit system. The ancilla is shown on top, and the controlled reversal gates are the pair of $C_Y$ gates. The phase gate $P(2Et)$ multiplies a phase of $\exp(i2Et)$ to the $\ket{1}$ state of the ancilla and cancels the relative phase produced by the controlled time evolution for an eigenstate with energy $E$.}
\label{fig:1CycleCirc}
\end{figure}

Our search for energy eigenvalues consists of three separate ``energy scans," referring to a collection of cycles with the same resolution \(\sigma\) but varying target energy \(E\). Each subsequent scan has a finer resolution and smaller range. 
\begin{comment}The energy range for the first scan is chosen to be from $-5$ to $5$. These values are determined using the Gershgorin circle theorem \cite{gershgorin1931} together with some extra padding to allow for the tails of the probability distribution. 
\end{comment}
The purpose of our coarse-grained first scan ($\sigma=4$) is to locate all possible regions of the energy spectrum with energy eigenvalue peaks. We identify any region with a success probability well above the average background value of $\tfrac{1}{2^n}$ as a candidate for the second, more focused scan with finer resolution ($\sigma=14$).  
\begin{comment}Later in our discussion we give the expected statistical error $\epsilon_n(E)$ associated with our measurement of $P_n(E)$. 
\end{comment}
The second scan is used to more finely determine the structure of the peaks associated with each energy eigenvalue. The third and final scan uses the highest resolution ($\sigma=24$) to accurately determine the center of each eigenvalue peak. The final scan consists of about twenty circuits covering a narrow region around each peak identified in the second scan.

\subsection{IBM and Quantinuum results}
We have performed the rodeo algorithm using three and five cycles on the IBM Perth machine and five cycles on the Quantinuum H1-2 machine. For all cases, we follow the same procedure. The initial state is $\ket{\psi_I} = \ket{00}$.  The first energy scan is performed with $\sigma=4$, the second scan with $\sigma=14$, and the third and final scan with $\sigma=24$. In Table~\ref{tb:gate_counts}, we summarize the number of two-qubit entangling gates required with and without controlled gate networks.  The results we present were obtained using controlled gate networks.

\begin{table}[htb]
\begin{tabular}{|c|c|c|c|}
\hline
& \begin{tabular}[c]{@{}l@{}}IBM Perth\\ Three Cycles\end{tabular} & \begin{tabular}[c]{@{}l@{}}IBM Perth\\ Five Cycles\end{tabular} &  \begin{tabular}[c]{@{}l@{}}Quantinuum H1-2\\ Five Cycles\end{tabular}\\ \hline
With controlled gate network & 12 & 20 & 20 \\ \hline
Without controlled gate network & 60 & 100 & 100 \\ \hline
\end{tabular}
\caption{Number of two-qubit entangling gates with and without controlled gate networks. \label{tb:gate_counts}}
\end{table}

Let $N_c$ be the number of distinct circuits implemented in a given scan, each with a unique set of Gaussian random values for the times. Let $N_s$ be the number of shots used for each circuit. The observed success probability $P_n(E)$ for $n$ cycles equals the number of successful measurements of $0$'s for all $n$ cycles divided by $N_sN_c$. For all circuits performed in this study, we use $N_s=1024$. For the first scan, we take $N_c = 5$, for the second scan $N_c = 2$, and for the final scan $N_c=1$. Since we are drawing from a binomial distribution, we can estimate that the one-standard-deviation statistical error for our measurement of $P_n(E)$ will be $\epsilon_n(E)=\sqrt{P_n(E)[1-P_n(E)]/(N_sN_c)}$. Since the success probability has a peak at each energy eigenvalue, $E_k$, we can fit a Gaussian function to each peak in the final scan data and determine the eigenvalues $E_k$.

\begin{figure}[htb]
\centering
\subfigure[]{\includegraphics[width=8cm]{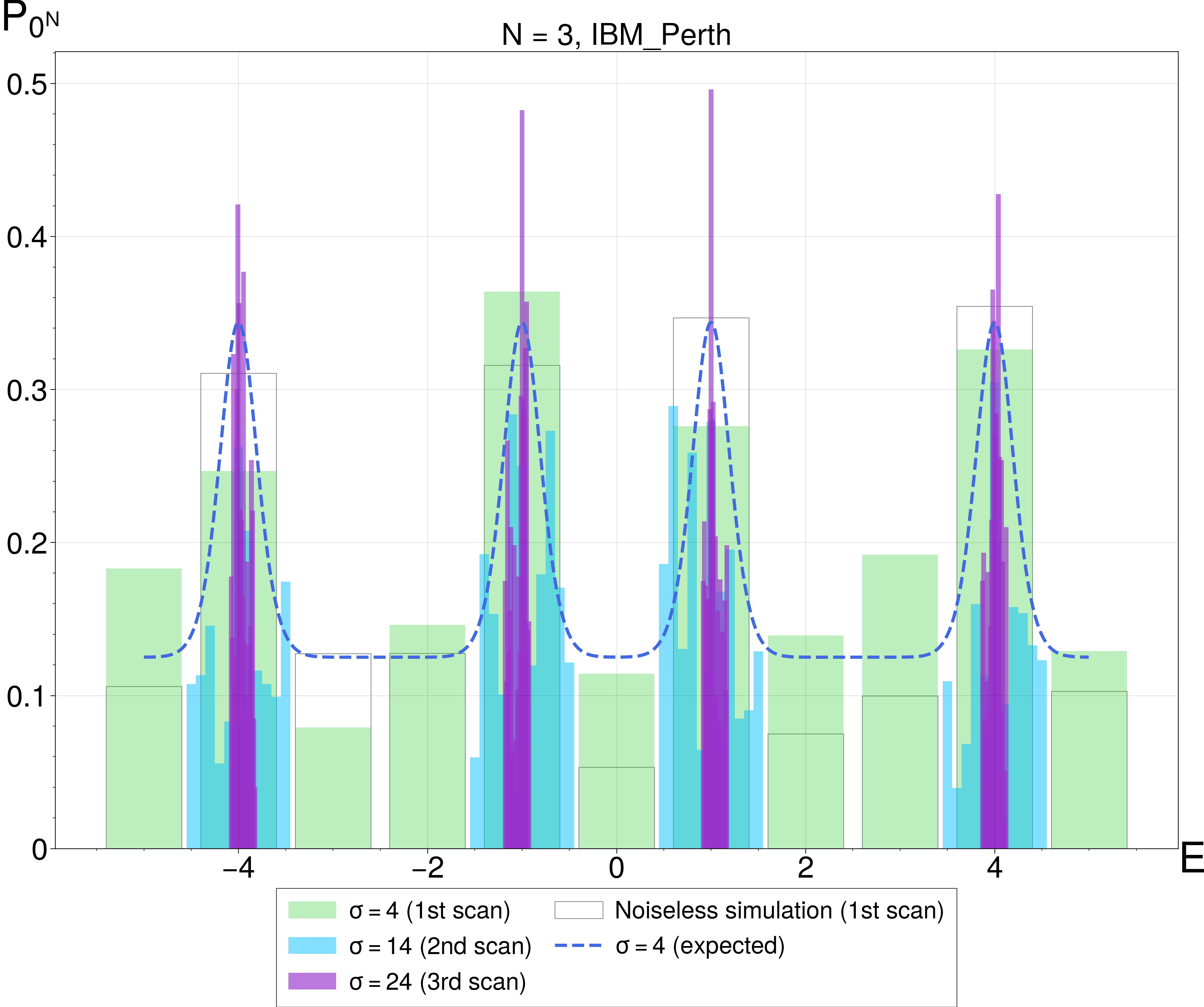}}
    \caption{Success probabilities corresponding to three scans of the rodeo algorithm for the IBM Perth system with three cycles.  The dashed line is the expected success probability from classical calculations of the first scan, and the white boxes are noiseless simulations at the energies of first scan.}
    \label{fig:three}
\end{figure}

The results for three cycles of the rodeo algorithm on IBM Perth are shown in the third column of \cref{tb:results}. The success probabilities versus energy are shown in \cref{fig:three}.  The curve with the dashed lines shows the exact analytical results for $\sigma = 4$. We also show the results for the first scan using a noiseless classical simulation. The peak success probabilities for the first scan are about $15\%$ lower than the values from the noiseless classical simulations. The average CNOT error for the portion of the IBM Perth device used is about $1\%$, and we expect that the largest source of noise is due to the 12 CNOT gates needed for each circuit. Compared with the exact results, the RMS deviation for the four eigenvalues is $0.010$, which is $0.12\%$ of the full range of the energy spectrum.  Meanwhile, the RMS value for the one-sigma uncertainties of the four eigenvalues is $0.006$ or $0.07\%$ of the full energy range.  We observe that the relative accuracy of the energy measurement is far better than the fidelity of the quantum circuits and is a striking example of the strong noise resilience of the rodeo algorithm.  

\begin{table}[htb]
\begin{tabular}{|c|c|c|c|c|}
\hline
& Exact & \begin{tabular}[c]{@{}l@{}}IBM Perth\\ Three Cycles\end{tabular} & \begin{tabular}[c]{@{}l@{}}IBM Perth\\ Five Cycles\end{tabular} &  \begin{tabular}[c]{@{}l@{}}Quantinuum H1-2\\ Five Cycles\end{tabular}\\ \hline
\,\(\ket{\psi_0}\)\, &\, -4\, &\, -4.0022(49)\, & \, -4.0006(42)\, &\, -3.9982(21)\, \\ \hline
\,\(\ket{\psi_1}\)\, &\, -1\, &\, -0.9829(56)\, & \,-0.9927(40)\, &\, -1.0083(39)\, \\ \hline
\,\(\ket{\psi_2}\)\, &\, 1\, &\, 1.0007(26)\, & \,1.0008(19)\, &\, 1.0028(22)\, \\ \hline
\,\(\ket{\psi_3}\)\, &\, 4\, &\, 4.0093(84)\, & \,3.9982(25)\, &\,  4.0036(17)\, \\ \hline
\end{tabular}
\caption{Energy eigenvalue estimates from the final pass of the rodeo algorithm with differing numbers of cycles on different devices. The exact values are provided for comparison. \label{tb:results}}
\end{table}

\begin{figure}[htb]
\centering
\subfigure[]{\includegraphics[width=8cm]{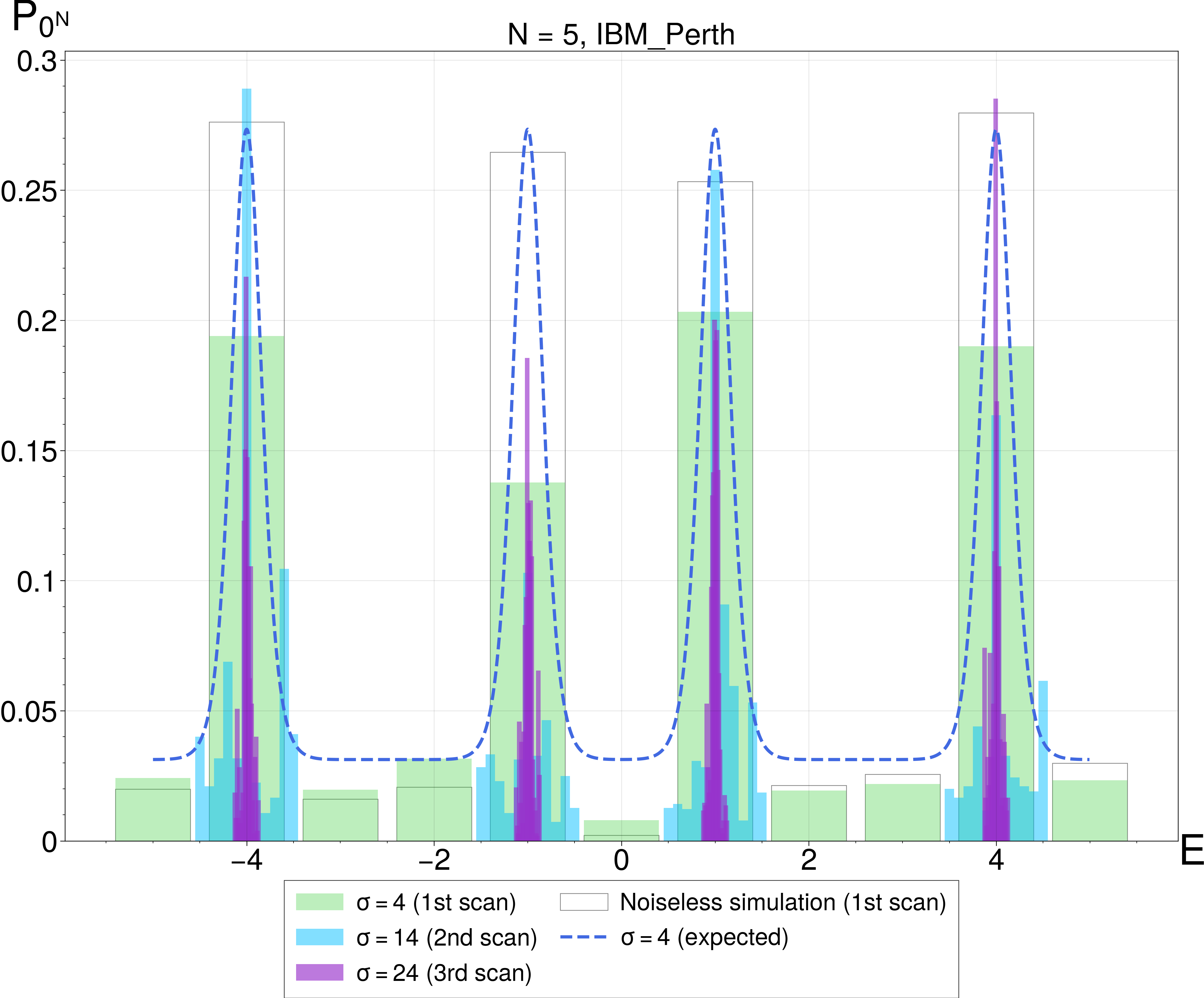}}
\subfigure[]{\includegraphics[width=8cm]{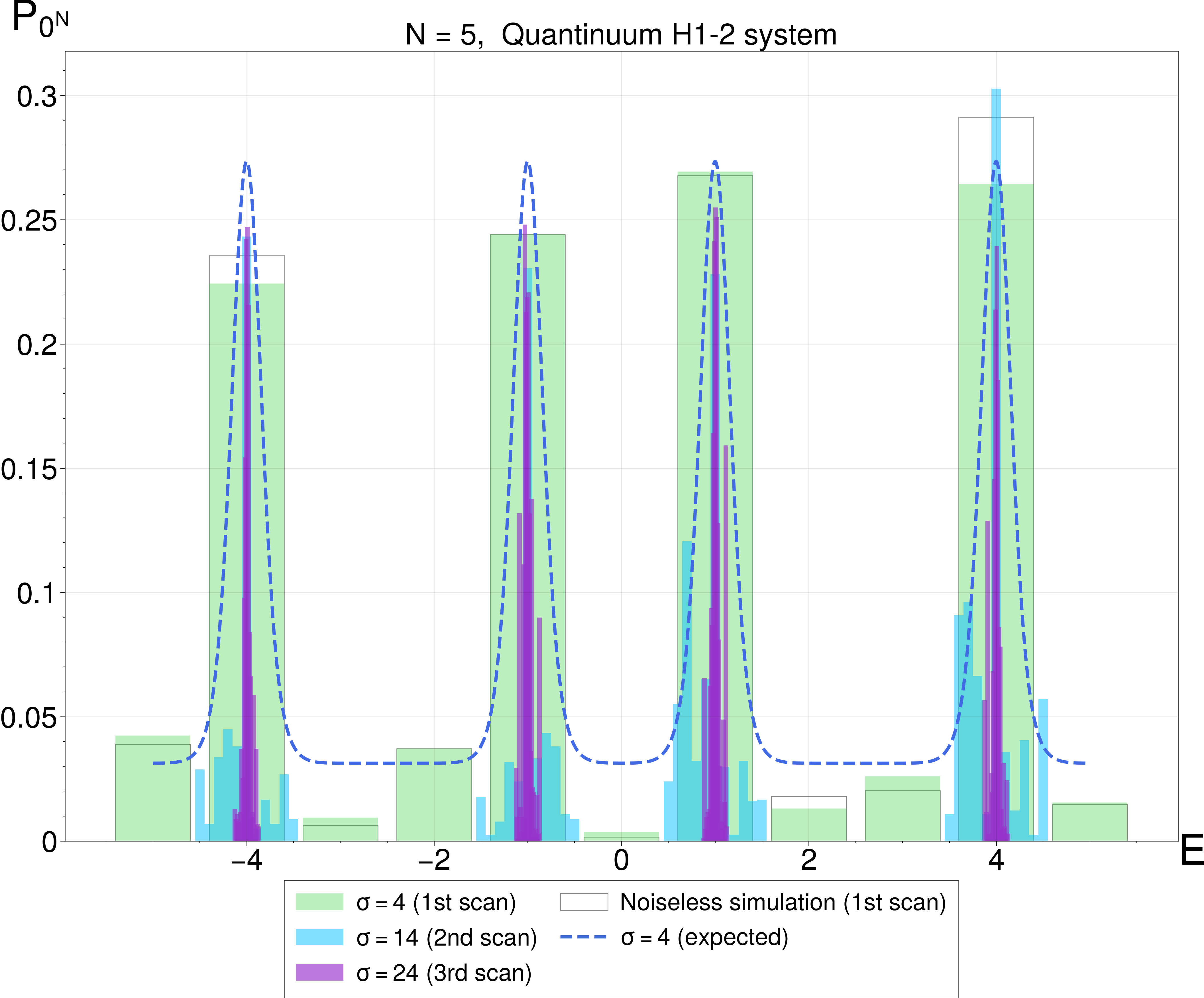}}
    \caption{Success probabilities corresponding to three scans of the rodeo algorithm for (a) the IBM Perth system with five cycles and (b) the Quantinuum H1-2 system with five cycles.  The dashed line is the expected success probability from classical calculations of the first scan, and the white boxes are noiseless simulations at the energies of first scan.}
    \label{fig:five}
\end{figure}

The success probability versus energy for five cycles on IBM Perth are shown in Panel (a) of \cref{fig:five}. In this case we see that the peak success probabilities for the first scan are about $30\%$ lower than the values from the noiseless classical simulations --- a significant reduction in performance. In this case we are using 20 CNOT gates. The results of the fit for five cycles on IBM Perth are shown in the fourth column of \cref{tb:results}. When compared with the exact results, the RMS deviation for the four eigenvalues is $0.004$, which is $0.05\%$ of the full range of the energy spectrum. The RMS value for the one-sigma uncertainties of the four eigenvalues is $0.003$ or $0.04\%$ of the full energy range.  We observe that the relative accuracy of the energy measurement is many orders of magnitude better than the roughly $30\%$ error rate for the quantum circuits.

The success probability versus energy for five cycles on the Quantinuum H1-2 are shown in Panel (b) of \cref{fig:five} \cite{H1-2}. In contrast with the IBM Perth results, the peak success probabilities for the first scan are close to those from the noiseless classical simulations, as one might expect due to the much better gate fidelities for the Quantinuum H1-2. The error rate for the $R_{ZZ}$ gate at the time of the experiment was $0.3\%$, and the error rate for the $20$ $R_{ZZ}$ gates needed for five cycles explains the approximately $5\%$ reduction in peak heights compared with noiseless simulations. The results of the fit for five cycles on the Quantinuum H1-2 are shown in the fifth column of \cref{tb:results}. When compared with the exact results, the RMS deviation for the four eigenvalues is $0.005$, which is $0.06\%$ of the full range of the energy spectrum.  The RMS value for the one-sigma uncertainties of the four eigenvalues is $0.003$ or $0.03\%$ of the full energy range.  We again see that the accuracy of the energy measurement is several orders of magnitude better than the roughly $5\%$ error rate for the quantum circuits.

The noise robustness of the rodeo algorithm comes from the fact that it implements a simple strategy of reducing the spectral weight of eigenstates with the wrong energy \cite{Qian:2021wya}.  While noise will reduce the spectral weight of the desired eigenstate, the associated eigenvalue peak will still be visible if the peak height is not reduced below the random background level.  Therefore, we expect the peak height to be reduced by noise, but the peak location should be unaffected.  This is clearly demonstrated in the results described above.  See the Supplemental Materials for a detailed analysis using a simple noise model.  If we had performed the same calculations without using controlled gate networks, we estimate that the error bars for the IBM Perth results for three cycles would be a factor of $2$ larger due to a factor of $2$ decrease in the signal to noise ratio.  The IBM Perth results for five cycles would be a factor of $3$ larger, and for Quantinuum H1-2 the error bars would be about a factor of $1.3$ larger.

\section{Applications to nuclear lattice simulations}

\subsection{Nuclear lattice simulations}
A useful starting point for simulations of nuclear many-body systems using lattice effective field theory is the four-component attractive Hubbard model \cite{Lee:2008fa,Lu:2018bat,Lahde:2019npb,Shen:2021kqr,Shen:2022bak,Elhatisari:2022zrb,Meissner:2023cvo,Giacalone:2024ixe,Giacalone:2024luz,Shen:2024qzi,Lee:2025req},
\begin{equation}
    H = -\frac{1}{2M} \sum_{j} \sum_{\langle \bm{r}, \bm{r}' \rangle}
    a_{j,\bm{r}}^\dagger a_{j,\bm{r}'} - C\sum_{j<j'} \sum_{\bm{r}} a_{j,\bm{r}}^\dagger a_{j,\bm{r}} a_{j',\bm{r}}^\dagger a_{j',\bm{r}}.
    \label{eq:Hlatt}
\end{equation}
The notation $\langle \bm{r}, \bm{r}' \rangle$ denotes nearest neighbors, and the index $j$ labels the four possible spin and isospin values for the nucleons.  We take the lattice to be a $L^3$ periodic cube, and we make a separate copies of the $L^3$ lattice for each index $j$.  The interaction between two nucleons with spin-isospin indices $j$ and $j'$ is implemented using a product of $Z$ operators for indices $j$ and $j'$ at each lattice site $\bm{r}$.  The antisymmetrization of the identical fermions with the same index $j$ can be handled using Jordan-Wigner strings.  The Jordan-Wigner strings for the hopping terms between $\bm{r}$ and $\bm{r}'$ can be organized in a manner that maximizes the cancellations of CNOT gates.  We also note some recent work in the literature on performing nuclear lattice simulations on quantum computers \cite{Roggero:2019myu,Baroni:2021xtl,Watson:2023oov,Spagnoli:2025xvk,Gu:2025yww}.  

The implementation of lattice simulations with many nucleons will be discussed in future work.  Here we focus on the implementation for a single nucleon with Hamiltonian,
\begin{equation}
    H = -\frac{1}{2M} \sum_{\langle \bm{r}, \bm{r}' \rangle}
    a_{\bm{r}}^\dagger a_{\bm{r}'}.
    \label{eq:Hfree}
\end{equation}
Since Fermi-Dirac statistics are not needed for a single nucleon, we represent the annihilation and creation operators on a quantum computer without including Jordan-Wigner strings.  In the following, we assume that $L$ is even so that we can partition the hopping terms $\langle \bm{r}, \bm{r}' \rangle$ into even and odd terms.  For each of the three spatial directions, we perform time evolution using Trotter steps that interleave the even and odd hopping terms.  This is illustrated in Fig.~\ref{fig:3Dhop}.

\begin{figure}
\centering
\resizebox{10.5cm}{!}{
\begin{quantikz}[row sep=0.3cm, column sep=0.22cm]
  %-------------------- q1
  \lstick{$q_1$}
    & \gate[2][1.5cm]{e^{i c (X_1X_2 + Y_1Y_2)}} \qw
    & \qw
    & \qw                          % short BEFORE-gap segment
    & \cdots \;          % (GAP column with centered dots)
    & \qw                          % short AFTER-gap segment
    & \gate[2][1.5cm]{e^{i c (X_1X_2 + Y_1Y_2)}} \qw
    & \qw
    & \qw \\
  %-------------------- q2
  \lstick{$q_2$}
    &                               % lower half of (12) box above
    & \gate[2][1.5cm]{e^{i c (X_2X_3 + Y_2Y_3)}} \qw
    & \qw                          % short BEFORE-gap segment
    & \cdots \;          % (GAP column with centered dots)
    & \qw                          % short AFTER-gap segment
    &                               % lower half of (12) box above
    & \gate[2][1.5cm]{e^{i c (X_2X_3 + Y_2Y_3)}} \qw
    & \qw \\
  %-------------------- q3
  \lstick{$q_3$}
    & \qw
    &                               % lower half of (23) box above
    & \qw                          % short BEFORE-gap segment
    & \cdots \;          % (GAP column with centered dots)
    & \qw                          % short AFTER-gap segment
    & \qw
    &                               % lower half of (23) box above
    & \qw
\end{quantikz}
}
\caption{Trotterized time evolution for even and odd hopping terms along one of the three spatial directions.}
\label{fig:3Dhop}
\end{figure}

We can implement $e^{ic(X_1X_2+Y_1Y_2)}$ with 2 CNOT gates, as shown in Fig.~\ref{fig:expblock}. For time evolution with $N_t$ Trotter steps for each of the three spatial directions, we have a total of $6L^3N_t$ CNOT gates.  

\subsection{Controlled time evolution}
In order to perform quantum phase estimation or an energy filtering method such as the rodeo algorithm, we need to implement controlled time evolution of the Hamiltonian, as shown in Fig.~\ref{fig:control3D}.  These controlled unitary operators are used to prepare linear combinations of quantum states with different durations of time evolution.  There are many ways to implement the controlled unitary operators, and in the following we compare four approaches. 

\begin{figure}
\centering
\begin{equation*}
\resizebox{8.5cm}{!}{%
\begin{quantikz}[row sep=0.3cm, column sep=0.2cm]
  \lstick{$q_1$} & \gate[2][1.5cm]{e^{i c (X_1X_2 + Y_1Y_2)}} & \qw \\
  \lstick{$q_2$} &                                   & \qw
\end{quantikz}
\;=\;
\begin{quantikz}[row sep=0.3cm, column sep=0.2cm]
\lstick{$q_1$} & \gate{H} & \ctrl{1} & \gate{R_y(-2c)} & \ctrl{1} & \gate{H}          & \qw \\
\lstick{$q_2$} & \gate{S} & \targ{}    & \gate{R_y(-2c)} & \targ{}    & \gate{S^{\dagger}} & \qw
\end{quantikz}
}
\end{equation*}
\caption{Implementation of $e^{ic(X_1X_2+Y_1Y_2)}$ using 2 CNOT gates.}
\label{fig:expblock}
\end{figure}

\begin{figure}
\centering
\resizebox{10.5cm}{!}{
\begin{quantikz}[row sep=0.3cm, column sep=0.22cm]
  %-------------------- ancilla q (controls each box)
  \lstick{$q$}
    & \ctrl{1}                     % controls e^{ic( X1X2 + Y1Y2 )}
    & \ctrl{2}                     % controls e^{ic( X2X3 + Y2Y3 )}
    & [0.18cm]\qw                  % short segment BEFORE gap
    & [0.70cm]\kern-2em \cdots \;  % (GAP column with centered dots)
    & [0.18cm]\qw                  % short segment AFTER gap
    & \ctrl{1}                     % controls e^{ic( X1X2 + Y1Y2 )}
    & \ctrl{2}                     % controls e^{ic( X2X3 + Y2Y3 )}
    & \qw \\
  %-------------------- q1
  \lstick{$q_1$}
    & \gate[2][1.5cm]{e^{i c (X_1X_2 + Y_1Y_2)}} \qw
    & \qw
    & \qw                          % short BEFORE-gap segment
    & \kern-2em \cdots \;          % (GAP column with centered dots)
    & \qw                          % short AFTER-gap segment
    & \gate[2][1.5cm]{e^{i c (X_1X_2 + Y_1Y_2)}} \qw
    & \qw
    & \qw \\
  %-------------------- q2
  \lstick{$q_2$}
    &                               % lower half of (12) box above
    & \gate[2][1.5cm]{e^{i c (X_2X_3 + Y_2Y_3)}} \qw
    & \qw                          % short BEFORE-gap segment
    & \kern-2em \cdots \;          % (GAP column with centered dots)
    & \qw                          % short AFTER-gap segment
    &                               % lower half of (12) box above
    & \gate[2][1.5cm]{e^{i c (X_2X_3 + Y_2Y_3)}} \qw
    & \qw \\
  %-------------------- q3
  \lstick{$q_3$}
    & \qw
    &                               % lower half of (23) box above
    & \qw                          % short BEFORE-gap segment
    & \kern-2em \cdots \;          % (GAP column with centered dots)
    & \qw                          % short AFTER-gap segment
    & \qw
    &                               % lower half of (23) box above
    & \qw
\end{quantikz}
}
\caption{Controlled time evolution of the Hamiltonian.}
\label{fig:control3D}
\end{figure}

The first approach is Method A, which uses the ancilla qubit $q$ to control each of the gates in the circuit for $e^{i c (X_1X_2 + Y_1Y_2)}$.  This is shown in Fig.~\ref{fig:Method_A}.  We need 6 CNOTs for each of the Toffoli gates and 2 CNOTs for each of the controlled $R_y$ rotations, 2 CNOTs each for the controlled $S$ and $S^\dagger$ gates, and 1 CNOT for each of the controlled $H$ gates.  Although some CNOT cancellations are possible, this is clearly not a good starting point with a total of $22$ CNOT gates. For time evolution with $N_t$ Trotter steps for each of the three spatial directions, this corresponds to $66L^3N_t$ CNOT gates, which is $11$ times greater than what is needed for time evolution without controlled operations.  The main problem with this approach is that there is no awareness of the underlying structure of $e^{i c (X_1X_2 + Y_1Y_2)}$, which would allow fewer gates to be controlled by the ancilla qubit. 

\begin{figure}[ht]
\centering
\begin{equation*}
\resizebox{12cm}{!}{%
\begin{quantikz}[row sep=0.3cm, column sep=0.2cm]
  \lstick{$q$}   & \ctrl{1} & \qw \\
  \lstick{$q_1$} & \gate[2][1.5cm]{e^{i c (X_1X_2 + Y_1Y_2)}} & \qw \\
  \lstick{$q_2$} &                                   & \qw
\end{quantikz}
\;=\;
\begin{quantikz}[row sep=0.3cm, column sep=0.1cm]
  \lstick{$q$}   & \ctrl{1} & \ctrl{2} & \ctrl{2} & \ctrl{1} & \ctrl{2} & \ctrl{2} & \ctrl{1} & \ctrl{2} & \qw \\
  \lstick{$q_1$} & \gate{H} & \qw & \ctrl{1} & \gate{R_y(-2c)} & \qw & \ctrl{1} & \gate{H} & \qw & \qw \\
  \lstick{$q_2$} & \qw & \gate{S} & \targ{} & \qw & \gate{R_y(-2c)} & \targ{} & \qw & \gate{S^\dagger} & \qw
\end{quantikz}
}
\end{equation*}
\caption{Drawing of Method A, where the ancilla qubit $q$ controls each of the gates in the original circuit for $e^{i c (X_1X_2 + Y_1Y_2)}$.}
\label{fig:Method_A}
\end{figure}

The second approach is Method B, which uses the ancilla qubit $q$ to control each of the single qubit gates in the circuit for $e^{i c (X_1X_2 + Y_1Y_2)}$ that depend on $c$.  This is shown in Fig.~\ref{fig:Method_B}.  This approach is similar to the controlled gate network used in our variational subspace example.  We need 2 CNOTs each for the controlled $R_y$ rotations that depend on $c$, and so this gives $6$ CNOT gates.  For time evolution with $N_t$ Trotter steps for each of the three spatial directions, we have a total of $18L^3N_t$ CNOT gates, which is a factor of $3$ times greater than that for time evolution without controlling operations. This approach exploits the structure of $e^{i c (X_1X_2 + Y_1Y_2)}$ so that fewer gates need to be controlled by the ancilla qubit.

\begin{figure}[ht]
\centering
\begin{equation*}
\resizebox{10.5cm}{!}{%
\begin{quantikz}[row sep=0.3cm, column sep=0.2cm]
  \lstick{$q$}   & \ctrl{1} & \qw \\
  \lstick{$q_1$} & \gate[2][1.5cm]{e^{i c (X_1X_2 + Y_1Y_2)}} & \qw \\
  \lstick{$q_2$} &                                   & \qw
\end{quantikz}
\;=\;
\begin{quantikz}[row sep=0.3cm, column sep=0.2cm]
  \lstick{$q$}   & \qw & \qw & \ctrl{1} & \ctrl{2} & \qw & \qw & \qw \\
  \lstick{$q_1$} & \gate{H} & \ctrl{1} & \gate{R_y(-2c)} & \qw & \ctrl{1} & \gate{H} & \qw \\
  \lstick{$q_2$} & \gate{S} & \targ{} & \qw & \gate{R_y(-2c)} & \targ{} & \gate{S^\dagger} & \qw
\end{quantikz}
}
\end{equation*}
\caption{Drawing of Method B, which uses the ancilla qubit $q$ to control each of the single qubit gates in the circuit for $e^{i c (X_1X_2 + Y_1Y_2)}$.}
\label{fig:Method_B}
\end{figure}

The third approach is Method C, which uses controlled $Z$ gates on one of the two qubits as a controlled reversal gate. This is shown in Fig.~\ref{fig:Method_C}.  We need 1 CNOT for each controlled $Z$ gate, which corresponds to $4$ CNOT gates.  For time evolution with $N_t$ Trotter steps for each of the three spatial directions, we have a total of $6L^3N_t$ CNOT gates, which is the same as for time evolution without controlling operations.  We note that we are saving a factor of $2$ because the controlled reversal gates reduce the number of required steps to $N_t/2$ for the forward and backward directions.  We assume that $N_t$ is chosen to be even. 

\begin{figure}[ht]
\centering
\begin{equation*}
\resizebox{10cm}{!}{%
\begin{quantikz}[row sep=0.3cm, column sep=0.2cm]
  \lstick{$q$}   & \octrl{1} & \qw & \ctrl{1} & \qw \\
  \lstick{$q_1$} & \gate[2][1.5cm]{e^{i c (X_1X_2 + Y_1Y_2)}} & \qw & \gate[2][1.5cm]{e^{-i c (X_1X_2 + Y_1Y_2)}} & \qw \\
  \lstick{$q_2$} &                                   & \qw &                                   & \qw
\end{quantikz}
\;=\;
\begin{quantikz}[row sep=0.3cm, column sep=0.2cm]
  \lstick{$q$}   & \ctrl{1} & \qw & \ctrl{1} & \qw \\
  \lstick{$q_1$} & \gate{Z} & \gate[2][1.5cm]{e^{i c (X_1X_2 + Y_1Y_2)}} & \gate{Z} & \qw \\
  \lstick{$q_2$} & \qw      &                    & \qw      & \qw
\end{quantikz}
}
\end{equation*}
\caption{Drawing of Method C, which uses controlled $Z$ gates on one of the two qubits as a controlled reversal gate.}
\label{fig:Method_C}
\end{figure}

The fourth approach is Method D, which uses controlled $Z$ gates on the even or odd qubits of our bipartite lattice as controlled reversal gates.  The controlled reversal gates appear at the beginning and end of the Trotter product.  For time evolution with $N_t$ Trotter steps for each of the three spatial directions, we have a total of $3L^3N_t + L^3$ CNOT gates.  We note that this is about half the number of CNOT gates needed for time evolution without control by the ancilla qubit.  The comparison of the four different approaches, Methods A, B, C, and D, demonstrates rather clearly the utility of controlled gate networks to design efficient quantum circuits.

\begin{figure}
\centering
\resizebox{12cm}{!}{
\begin{quantikz}[row sep=0.3cm, column sep=0.22cm]
  %-------------------- ancilla q (continuous across the gap)
  \lstick{$q$}
    & \ctrl{1}                     % col 0: CZ(q→q1) start
    & \ctrl{3}                     % col 1: CZ(q→q3) start
    & \qw                          % col 2
    & \qw                          % col 3
    & [0.18cm]\qw                  % col 4 (short before gap)
    & [0.70cm]\qw                  % col 5 (GAP column; q stays continuous)
    & [0.18cm]\qw                  % col 6 (short after gap)
    & \qw                          % col 7
    & \qw                          % col 8
    & \ctrl{3}                     % col 9: CZ(q→q3) end
    & \ctrl{1}                     % col10: CZ(q→q1) end
    & \qw \\                       % col11
  %-------------------- q1
  \lstick{$q_1$}
    & \gate{Z}                     % col 0 (partner of \ctrl{1})
    & \qw                          % col 1
    & \gate[2][1.5cm]{e^{i c (X_1X_2 + Y_1Y_2)}} \qw  % col 2
    & \qw                          % col 3
    & \qw                          % col 4 (short before gap)
    & \kern-2em \cdots \;           % col 5 (GAP column with centered dots)
    & \qw                          % col 6 (short after gap)
    & \gate[2][1.5cm]{e^{i c (X_1X_2 + Y_1Y_2)}} \qw  % col 7
    & \qw                          % col 8
    & \qw                          % col 9
    & \gate{Z}                     % col10 (partner of final \ctrl{1})
    & \qw \\                       % col11
  %-------------------- q2
  \lstick{$q_2$}
    & \qw                          % col 0
    & \qw                          % col 1
    &                               % col 2 (lower half of (12) above)
    & \gate[2][1.5cm]{e^{i c (X_2X_3 + Y_2Y_3)}} \qw  % col 3
    & \qw                          % col 4 (short before gap)
    & \kern-2em \cdots \;           % col 5 (GAP column with centered dots)
    & \qw                          % col 6 (short after gap)
    &                               % col 7 (lower half of (12) above)
    & \gate[2][1.5cm]{e^{i c (X_2X_3 + Y_2Y_3)}} \qw  % col 8
    & \qw                          % col 9
    & \qw                          % col10
    & \qw \\                       % col11
  %-------------------- q3
  \lstick{$q_3$}
    & \qw                          % col 0
    & \gate{Z}                     % col 1 (partner of \ctrl{3})
    & \qw                          % col 2
    &                               % col 3 (lower half of (23) above)
    & \qw                          % col 4 (short before gap)
    & \kern-2em \cdots \;           % col 5 (GAP column with centered dots)
    & \qw                          % col 6 (short after gap)
    & \qw                          % col 7
    &                               % col 8 (lower half of (23) above)
    & \gate{Z}                     % col 9 (partner of final \ctrl{3})
    & \qw                          % col10
    & \qw                          % col11
\end{quantikz}
}
\caption{Drawing of Method D, which uses controlled $Z$ gates on the even or odd qubits of our bipartite lattice as controlled reversal gates.}
\label{fig:Method_D}
\end{figure}

\section{Discussion and outlook}

We have introduced a new scheme for quantum circuit design called controlled gate networks.  Rather than reducing the complexity of individual unitary operations, the goal is to toggle between the required unitary operations with the fewest number of CNOT gates.  We have discussed the general theory of controlled reversal gates and show that linear combinations of unitary operators can be implemented with fewer CNOT gates using controlled gate networks when the unitary operators have similar structures.

We then demonstrated the new approach with three examples. The first is a variational subspace calculation using QAOA for a two-qubit system.  As measured by the Qiskit transpiler, the use of controlled gate networks produces a factor of $5$ reduction in the number of CNOT gates and about a factor $4$ in the number of single qubit gates.  While there are more powerful transpiling software tools available, our use of the Qiskit transpiler for both the standard approach and the controlled gate network approach gives some measure of the relative efficiencies of the two approaches.

In our second example, we have demonstrated that with controlled reversal gates and $n=3$ cycles of the rodeo algorithm and $\sigma = 24$, we can resolve the energy eigenvalue with an error of about $0.010$ on IBM Perth, or $0.42(\sqrt{n}\sigma)^{-1}$. With $n=5$ cycles of the rodeo algorithm and $\sigma = 24$, we can resolve the energy eigenvalue with an error of about $0.005$ on IBM Perth, or $0.27(\sqrt{n}\sigma)^{-1}$; on Quantinuum H1-2, we can resolve the energy eigenvalue with an error of about $0.006$, or $0.32(\sqrt{n}\sigma)^{-1}$. 

The third example is the implementation of the controlled time evolution of a single nucleon on a three-dimensional lattice.  This is an important first step towards performing nuclear lattice simulations on a quantum computer.  We have explored four different methods for implementing the controlled time evolution, starting from a simple approach that does not exploit the structure of the operators and ending with an efficient technique using controlled reversal gates.  We find that the use of controlled reversal gates can reduce the number of CNOT gates by a substantial factor.  These gains in efficiency are very important for future calculations of nuclear lattice simulations that include many nucleons. In future work, we will show how a controlled gate network can also reduce the number of CNOT gates needed for the lattice interactions and antisymmetrization.  

Controlled gate networks represent a new paradigm for quantum circuit design, where controlling qubits are used to efficiently transform from one unitary operator to another unitary operator with a similar structure.  These controlled unitary operations are common in quantum state preparation algorithms, variational subspace calculations, and any application using linear combinations of unitary operators.  The approach appears to be quite powerful and general with many applications for quantum circuit design.

In addition to introducing controlled gate networks, we have also demonstrated that the rodeo algorithm with controlled reversal gates is a practical and efficient method for eigenvalue estimation on current quantum devices.  In order to accurately determine energy eigenvalues, the rodeo algorithm needs only to be able to distinguish the peak associated with an energy eigenvalue above the background level of $\tfrac{1}{2^n}$. 
Once this threshold criterion is reached, the accuracy of the rodeo algorithm to determine the energy eigenvalue is set by the sharpness of the energy eigenvalue peaks in Eq.~(\ref{eq:probability1}). The width of the peak scales as $(\sqrt{n}\sigma)^{-1}$, where $n$ is the number of cycles and $\sigma$ is the standard deviation of the Gaussian random times. 

\paragraph*{Acknowledgements}
We are grateful for discussions with Joey Bonitati, Eli Chertkov, Michael Foss-Feig, Gabriel Given, David Hayes, Ashe Hicks, Brian Neyenhuis, and Danny Samuel. We also thank Jay Gambetta and IBM Q for the research accounts and access to hardware. We acknowledge financial support from the U.S. Department of Energy through grants DE-SC0023658, DE-SC0021152, DE-SC0013365, DE-SC0024586, DE-SC0023175, DE-SC0026198, and the U.S. National Science Foundation through the grant PHY-2310620 and the Graduate Research Fellowship Program through the grant DGE-1848739.  This research used resources of the Oak Ridge Leadership Computing Facility, which is a DOE Office of Science User Facility supported under Contract DE-AC05-00OR22725.

\clearpage

\section*{Supplemental Materials}

\subsection*{Success probabilities}
For each rodeo cycle, the random times $t$ are selected according to the normalized Gaussian distribution 
\begin{equation}
p(t) = \tfrac{1}{\sqrt{2\pi}\sigma}e^{-\tfrac{t^2}{2\sigma^2}}.
\end{equation}  
If our initial is an eigenstate with energy $E_{k}$, then the success probability for one cycle with target energy $E$ is
\begin{equation}
  \frac{1}{\sqrt{2\pi}\sigma}\int dt e^{-\frac{t^2}{2\sigma^2}} \cos^2 [(E_{k}-E)\tfrac{t}{2}] =  \frac{1+e^{-(E_{k}-E)^2\sigma^2/2} }{2} \label{eq:continuous},
\end{equation}
and the probability of success for all $n$ cycles is 
\begin{equation}
    \left[ \frac{1+e^{-(E_{k}-E)^2\sigma^2/2} }{2} \right]^n.
\end{equation}
Let us now consider an initial state $\ket{\psi_I}$, which we can decompose into energy eigenstates
\begin{equation}
  \ket{\psi_I} = \sum_k \braket{E_k|\psi_I}\ket{E_k}. 
\end{equation}
With this initial state, the success probability after $n$ rodeo cycles will then equal
\begin{equation}
    P_n(E) = \sum_k  \frac{\left[1+e^{-(E_k-E)^2\sigma^2/2}\right]^n \left| \braket{E_k|\psi_I} \right|^2 }{2^n}.
    \label{eq:probability2}
\end{equation}

\subsection*{Noisy Hamiltonian evolution}
Consider a simple noise model in which the circuit still implements Hamiltonian time evolution for each cycle, but noise in the Hamiltonian shifts the energy eigenvalues $E_k$ by some amount $\delta E_{k,i}$ for the $i$th cycle of the algorithm, where $\delta E_{k,i}$ is random shot-to-shot and is independently distributed between cycles. To first order in the noise, we assume that the eigenvectors $|E_k\rangle$ are unchanged. The noisy success probability can be obtained by averaging $P_n(E)$ over the distribution $p(\delta E_{k,n})$ of the noise:
\begin{equation}
P_{n,\text{noisy}}(E) = \sum_k \frac{|\langle E_k | \psi_I\rangle|^2 }{2^n} \prod_{i=1}^n \int d(\delta E_{k,i}) p(\delta E_{k,i}) \left[ 1 + e^{-(E_k + \delta E_{k,i} - E)^2 \sigma^2 / 2} \right] \,.
\end{equation}
In the simple case where $\delta E_{k,i}$ is Gaussian distributed with mean 0 and standard deviation $\varepsilon_k$, then $p(\delta E_{k,i}) = \frac{1}{\sqrt{2 \pi \varepsilon_k^2}} e^{-(\delta E_k)^2/(2 \varepsilon_k^2)}$ and
\begin{equation}
    \int d(\delta E_{k,i}) p(\delta E_{k,i}) \left[ 1 + e^{-(E_k + \delta E_{k,i} - E)^2 \sigma^2 / 2} \right] = 1 + \frac{e^{-(E_k - E)^2 \sigma^2 / [2 (1 + \varepsilon_k^2 \sigma^2)]}}{\sqrt{1 + \varepsilon_k^2 \sigma^2}} \,.
\end{equation}
The centers of the peaks thereby remain at $E_k$. When the eigenenergies are well-separated, the peak height at $E = E_k$ is reduced by a factor of $\left[1 + \frac{1}{\sqrt{1 + \varepsilon_k^2 \sigma^2}}\right]^n / 2^n$, which gives exponential suppression in the number $n$ of cycles. When $\varepsilon_k^2 \sigma_k^2 \ll 1$, one obtains
\begin{equation}
P_{n,\text{noisy}}(E_k) \approx |\langle E_k | \psi_I\rangle|^2 \left[1 - \varepsilon_k^2 \sigma^2 / 4 \right]^n \,.
\end{equation}

One can also consider the case when $\delta E_{k,i} = \delta E_k$ is independent shot-to-shot, but the same for all cycles. This might, for instance, reflect slowly-varying coherent errors arising from calibration inaccuracies. Assuming sufficient sampling that averages over the error distribution,
\begin{equation}
    P_{n,\text{noisy}}(E) = \sum_k \frac{|\langle E_k | \psi_I\rangle|^2 }{2^n} \int d(\delta E_{k}) p(\delta E_{k}) \left[ 1 + e^{-(E_k + \delta E_{k} - E)^2 \sigma^2 / 2} \right]^n \,.
\end{equation}
For a Gaussian error distribution, this will again leave the centers of the peaks unchanged. To find the suppression of the peak height, one can set $E = E_k$ and perform the binomial expansion and the integration. Assuming the peaks are well-separated, one obtains
\begin{equation}
    P_{n,\text{noisy}}(E_k) \approx \frac{|\langle E_k | \psi_I\rangle|^2 }{2^n} \sum_m \binom{n}{m} \frac{1}{\sqrt{1 + m \varepsilon^2 \sigma^2}} \,.
\end{equation}
Assuming $m \varepsilon_k^2 \sigma^2 \ll 1$, then to lowest order one obtains
\begin{equation}
P_{n,\text{noisy}}(E_k) \approx |\langle E_k | \psi_I\rangle|^2 \left[ 1 - n \varepsilon_k^2 \sigma^2 / 4 \right] \,.
\end{equation}
In this case, one sees only a linear decay of the peak height in the number of rodeo cycles.

\bibliography{References}

\end{document}